\documentclass[aps,prb,reprint,twocolumn,citeautoscript,superscriptaddress,showkeys,showpacs]{revtex4-1}
\usepackage{xr}
\usepackage[fleqn,tbtags]{amsmath}
\usepackage{amsfonts, amssymb,amsxtra,textcomp}
\usepackage[]{graphicx}
\usepackage{grffile}
\usepackage{bbm}
\usepackage{bm}
\usepackage{color}
\usepackage{bbold}
\usepackage{graphicx}
\usepackage[colorlinks=true,citecolor=blue,linkcolor=blue]{hyperref}

\allowdisplaybreaks

\usepackage{color}

\newcommand{\be}{\begin{equation}}
\newcommand{\ee}{\end{equation}}
\newcommand{\bea}{\begin{eqnarray}}
\newcommand{\eea}{\end{eqnarray}}
\newcommand{\bse}{\begin{subequations}}
\newcommand{\ese}{\end{subequations}}

\usepackage{color}                                      
\definecolor{d_red}{cmyk}{0.00, 0.81, 1.00, 0.27}
\definecolor{d_orange}{cmyk}{0.00, 0.33, 1.00, 0.00}
\definecolor{d_blue}{cmyk}{0.78, 0.47, 0.00, 0.20}
\definecolor{d_lgreen}{cmyk}{0.07, 0.00, 0.79, 0.29}
\definecolor{d_green}{cmyk}{0.66, 0.00, 0.71, 0.56}
\definecolor{d_blue}{cmyk}{0.78, 0.47, 0.00, 0.20}
\definecolor{d_dblue}{cmyk}{0.91, 0.79, 0.00, 0.22}
\definecolor{d_pink}{cmyk}{0.0, 0.79, 0.37, 0.29}
\definecolor{d_purple}{cmyk}{0.16, 0.54, 0.00, 0.70}
\definecolor{d_paleblue}{cmyk}{0.669, 0.338, 0.00, 0.373}
\definecolor{d_dpaleblue}{cmyk}{0.441, 0.290, 0.00, 0.580}
\definecolor{d_brown}{cmyk}{0.0, 0.490, 0.930, 0.350}
\definecolor{d_turquoise}{cmyk}{0.630, 0.04, 0.0, 0.440}
\definecolor{KIT-green}{RGB}{0, 150,130}
\definecolor{KIT-blue}{RGB}{70,100,170}

\def\bmx{\begin{pmatrix}}
\def\emx{\end{pmatrix}}

\usepackage{hyperref}
\usepackage[figure,table]{hypcap}               
\hypersetup{
pdftitle = {},
pdfsubject = {},
pdfauthor = {},
pdfkeywords = {},
pdfcreator = {},
pdfproducer = {LaTeX with hyperref},
colorlinks = true,
linkcolor = blue,
anchorcolor = blue,
citecolor = blue,
filecolor = red,
menucolor = red,
pagecolor = red,
urlcolor  = blue,
breaklinks = true,
pdfstartview = FitV,
pdfhighlight = /I,
pdfpagelayout = TwoColumn,
hypertexnames=true
}

\begin{document}

\title{Hydrodynamic transport in the Luttinger-Abrikosov-Beneslavskii non-Fermi liquid}

\author{Julia M. Link}
\affiliation{Department of Physics, Simon Fraser University, Burnaby, British Columbia V5A 1S6, Canada}

\author{Igor F. Herbut}
\affiliation{Department of Physics, Simon Fraser University, Burnaby, British Columbia V5A 1S6, Canada}

\date{\today}
\begin{abstract}

We determine the shear viscosity and the dc electrical conductivity of interacting three-dimensional Luttinger semimetals, which have a quadratic band touching point in the energy spectrum, in the hydrodynamic regime. It is well known that when the chemical potential is right at the band touching point the long-range Coulomb interaction induces the Luttinger-Abrikosov-Beneslavskii (LAB) phase at $T=0$, which is an interacting, scale-invariant, non-Fermi-liquid state of electrons. Upon combining the renormalization-group (RG) analysis near the upper critical spatial dimension of 4 with the Boltzmann kinetic equation, we determine the universal ratio of viscosity over entropy, and the electrical dc conductivity of the system at the interacting LAB fixed point of the RG flow. The projection of the Coulomb interaction on the eigenstates of the system is found to play an important quantitative role for the scattering amplitude in the collision integral, and the so-called Auger-processes make a large numerical contribution to the inverse scattering time in
 the transport quantities. The obtained leading-order result suggests that the universal ratio of the viscosity over entropy, when extrapolated to the physical three spatial dimensions, is above, but could be rather close to, the Kovtun-Son-Starinets lower bound.
\end{abstract}

\maketitle

\section{Introduction}

When the influence of the electron-phonon interaction and impurities can be neglected and the physics assumed to be governed by the collisions of the quasiparticles due to their mutual interactions, the universal features of the hydrodynamic regime become manifest \cite{Damle1997}. One important transport property in this regime is the shear viscosity, which measures dissipative effects due to internal friction in the fluid. The magnitude of the shear viscosity determines whether the flow of quasiparticles in a crystal is laminar or turbulent. The latter, for example, leads to the tendency of the system to form vortices, as is the case for the flow of electrons and holes in graphene \cite{Levitov2016}. The true measure of viscosity is the celebrated Kovtun-Son-Starinets (KSS) ratio of shear viscosity $\eta$ and entropy $s$ \cite{Kovtun2005}, for which the lower bound in the form
\begin{equation}
 \frac{\eta}{s} \ge \frac{\hbar}{4\pi k_B}
 \:,
 \label{eq:viscoverentropy}
\end{equation}
was derived for a relativistic quantum system with the dynamical exponent $z=1$, using the correspondence between strongly interacting quantum and classical (gravitational) field theories. It was also later shown that gravitational theories that correspond to strongly interacting quantum systems with $z=2$ obey the same bound for the ratio of viscosity over entropy as defined in Eq.~\eqref{eq:viscoverentropy} \cite{Kovtun2009}. In physical terms, this lower bound can be understood as that the mean free scattering path of the quasiparticles should not be smaller than their thermal length. The closer a system comes to the lower bound, the more effectively strongly interacting it is. Such strongly interacting systems with a very small ratio of $\eta/s$ are therefore sometimes called ``nearly perfect fluids".

Of recent interest, for example,  was the hydrodynamic transport in graphene \cite{Narozhny2017,Lucas2018}, which is a prototypical two-dimensional material with two linear energy bands crossing at the Fermi point. Some of the features of graphene under scrutiny were the breakdown of the Wiedemann-Franz law \cite{Crossno2016}, the small ratio $\eta/s$ \cite{Mueller2009, Herbut2009}, the logarithmically diverging dc conductivity due the Coulomb interaction \cite{Fritz2008}, the negative local resistance \cite{Levitov2016,Bandurin2016}, and the giant magneto-drag \cite{Titov2013}. Further studies of two-dimensional materials with an anisotropic band touching point (parabolic dispersion in one direction and a linear one in the perpendicular direction) found fascinating hydrodynamic transport behavior such as anisotropy in the electrical conductivity, and a possible modification of the lower bound \cite{Link2018}.

In this paper, we extend the study of the electronic transport of systems with an energy-band-touching point in the hydrodynamic regime to the important and ubiquitous class of three-dimensional systems where two energy bands cross  quadratically at the Fermi level, which we will call ``Luttinger semimetals" \cite{Luttinger}. Many compounds, such as gray tin, HgTe, irridates, or half-Heuslers, fall into this general class. It is well known that Coulomb electron-electron interaction in this situation is not screened, and its long-range nature causes the Luttinger semimetals  to adopt the scale-invariant non-Fermi-liquid ground state \cite{Abrikosov1974}. Here we determine the shear viscosity $\eta$ and the dc electrical conductivity $\sigma$ of the interacting Luttinger semimetals, and in particular, the influence of the long-range Coulomb interaction on these transport quantities in the hydrodynamic regime. We combine the renormalization-group (RG) analysis in $\epsilon=4-d$ expansion, where $d$ is the spatial dimension of the system, with the Boltzmann formalism for the transport. The projection of the Coulomb interaction on the eigenstates of the system is found to play an important role in the determination of the transport quantities. Furthermore, the so-called Auger processes, \textit{i.e.}, scattering processes where two holes (electrons) scatter into a particle-hole state and vice versa, significantly contribute to the scattering time that enters both the conductivity and the viscosity. As a result of the somewhat long calculation, presented in detail in the Appendixes, we find that the interacting Luttinger semimetals exhibit a power-law temperature dependence of the dc conductivity, and that the Coulomb interaction reduces the electrical conductivity from the non-interacting value. Most importantly, the universal ratio of viscosity over entropy is determined to be
\begin{equation}
\frac{\eta}{s}=\frac{\hbar}{k_B} \frac{0.137}{\epsilon^2} +\mathcal{O}(1/\epsilon)
\:.
\label{eq:result_visc}
\end{equation}
This leading order result indicates that the three-dimensional ($\epsilon=1$) interacting Luttinger semimetals in their non-Fermi-liquid ground state may also deserve to be called nearly perfect fluids, and that they could be closer to the lower bound than many other strongly interacting systems \cite{Wilks1968,Lacey2007,Schaefer2009}. Of course, the extension of the leading-order result down to the physical case of $\epsilon=1$ may be questioned, as always. It will therefore be interesting to see if the formalism used here, or some alternative method such as large-$N$, the Kubo formula \cite{Bradlyn2012,Link2018a}, and the Mori-Zwanzig memory-matrix formalism\cite{Zaanen2015,Hartnoll2018}, which does not rely on the concept of quasiparticles, may be pushed further to get a better estimate of this interesting universal value. 

The rest of the paper is organized as follows: First the Luttinger Hamiltonian and the RG-analysis of the system leading to the Luttinger-Abrikosov-Beneslavskii (LAB) fixed point value for the coupling constant are introduced in Secs.~\ref{sec:LuttignerHam} and \ref{sec:LABfixedpoint}, respectively. In Sec.~\ref{sec:Boltzmann}, the Boltzmann formalism is implemented in order to determine the shear viscosity $\eta$ and the electrical conductivity $\sigma$ of the Luttinger semimetals. A special focus is set on the collision integral, where we demonstrate that the projection of the Coulomb interaction on the eigenstates of the system is important. The result of our calculation is presented in Sec.~\ref{sec:Results}, and a summary can be found in Sec.~\ref{sec:Summary}.

\section{Luttinger Hamiltonian}
\label{sec:LuttignerHam}
Luttinger semimetals have a band touching point that is protected by the symmetries of the cubic lattice ($T_h$ or $O_h$)\cite{Luttinger, ABRIKOSOV1996}. However, in contrast to graphene and the Weyl semimetals, the energy dispersion is not linear here but quadratic. The quadratic band crossing in a system with a cubic lattice symmetry is in general described by the Luttinger Hamiltonian with the three mass parameters:
\begin{eqnarray}
H &=&
\frac{1}{2m}\left[ \frac{5}{4} p^2-(\boldsymbol{p} \cdot \boldsymbol{J} )^2 \right]
 +\frac{p^2}{2 \tilde{M}_0} -\frac{p_x^2 J_x^2+p_y^2 J_y^2 + p_z^2 J_z^2}{2 M_c} \nonumber \\
&=&
\frac{1}{2m} \sum_{a=1}^5 d_a(\boldsymbol{p}) \gamma_a
+\frac{p^2}{2 \tilde{M}_0} +\frac{d_4(\boldsymbol{p}) \gamma_4 + d_5(\boldsymbol{p}) \gamma_5 }{2 M_c}\:,
\end{eqnarray}
where $\boldsymbol{J}$ are the three components of the $j=3/2$ total angular momentum. The Luttinger Hamiltonian can be rewritten in terms of the spherical harmonics for the angular momentum of 2 with $d_a(\boldsymbol{p})=\sqrt{d/[2(d-1)]} p_i \left( \Lambda^a \right)_{ij} p_j$, where $\Lambda^a$ are the five real Gell-Mann matrices defined in the Appendix~\ref{ap:theSystem},  and $\gamma_a$ are the Dirac matrices obeying the usual Clifford algebra $\{ \gamma_a, \gamma_b \}=2 \delta_{ab}$. For the purpose of our calculation, the Hamiltonian in the latter form is more useful, since it can be easily extended to $d=4$ upon rewriting the spherical harmonics and the $\gamma$-matrices in $d=4$, as explained in Appendix~\ref{ap:theSystem}  \cite{Janssen2015}.
The resulting energy dispersion is given by:
\begin{equation}
 \epsilon_{\boldsymbol{p},\pm}=\frac{p^2}{2 M_0} \pm
 \sqrt{\left( \frac{p^2}{2m} \right)^2+ \frac{m + 2 M_c}{4 m M_c^2} (\sum_i k_i^4-\sum_{i\neq j} k_i^2 k_j^2 )}
 \:.
\end{equation}
The first term in the Luttinger Hamiltonian inversely proportional to $m$ denotes the particle-hole and rotationally-symmetric parabolic energy dispersion, while the second and third terms introduce particle-hole and cubic symmetry breaking perturbations, respectively.

\section{LAB fixed point}
\label{sec:LABfixedpoint}
The systems of Luttinger fermions interacting solely via long-range Coulomb interaction is described by the Lagrangian
\begin{equation}
 \mathcal{L}=\psi^{\dagger} (\partial_{\tau} + i \phi + H) \psi +\frac{1}{ 8 \pi e^2} (\nabla \phi)^2
 \:.
\end{equation}
The scalar field $\phi$ mediates the long-range Coulomb interaction $ 4 \pi e^2 /q^2$, or in real space $\propto e^2/r^{d-2}$. The Coulomb interaction coupling constant $e^2$ becomes a relevant coupling at the non-interacting fixed point in $d<4$ dimensions, which calls for the introduction of a small parameter $\epsilon=4-d$ to formulate a perturbative approach. The quantum critical phase, the so-called Luttinger-Abrikosov-Beneslavskii (LAB) phase \cite{Abrikosov1974,Moon2013}, is then found from the flow equation of the coupling constant
\begin{equation}
\frac{d \tilde{e}^2}{d \log b}=\epsilon \tilde{e}^2 - \frac{3 N_d +1}{6 } \tilde{e} ^{4}
\:,
\label{eq:floweq}
\end{equation}
where we have introduced the dimensionless charge ${\tilde{e}^2= ( S_4/(2\pi)^4 ) 4 \pi e^2 \Lambda^{d-z-2}}$, with $S_d = 2 \pi^{d/2}/\Gamma(d/2)$ as the area of the sphere in $d$ dimensions. $\Lambda$ is the arbitrary momentum cutoff. The stable fixed point and the dynamical critical exponent $z$ are then, to the leading order in $\epsilon$,
\begin{equation}
 \tilde{e} _*^{2}=\frac{6}{3N_{d}+1}\epsilon
\:, \qquad
 z=2-\frac{1}{6}\tilde{e} _*^{2}
 \:,
 \label{eq:fixedpoint}
\end{equation}
where $N_d$ denotes the degeneracy of the energy bands of the Luttinger Hamiltonian when generalized to $d$ dimensions ($N_3=2$, $N_4=8$).
The terms introducing the particle-hole and the cubic asymmetries in the Hamiltonian are known to be irrelevant at the stable fixed point \cite{Abrikosov1974,Moon2013, Boettcher2017}  and will therefore be neglected in the following. We should mention that there is an ambiguity regarding how to implement the RG procedure in general dimension; here we extended first the Luttinger Hamiltonian to $d=4$,\cite{Abrikosov1974, Janssen2015} performed the angular momentum integration of 4-dimensional momenta, and then implemented dimensional regularization \cite{Herbut2007} on the absolute value of the momentum $p$. For this reason, the numbers appearing in the flow equation and the dynamical exponent differ from those in Ref.~\onlinecite{Moon2013}, for example, where only the last step was implemented, with the angular integrals performed in $d=3$.

In this work, we also neglect the possible effects of the short-range components of the Coulomb interaction on the LAB fixed point, which may ultimately cause its annihilation and the replacement with the runaway flow of the RG \cite{Herbut2014, Janssen2017}.  The rationale is that even if this should happen before the parameter $\epsilon$ reaches unity, there should be a large window of crossover scales at which the scaling appropriate to the LAB fixed point is still visible. In this window, the characteristic power-law behavior of the viscosity and the conductivity,  $\eta \sim T^{d/z}$ and $\sigma \sim T^{(d-2)/z}$, should be discernible. Neglecting the possible annihilation of the LAB fixed point simply means that these power-laws extend all the way to $T=0$. We will assume this here for the sake of simplicity.

To compute the numerical constants of the above transport coefficients, we next turn to the kinetic theory.

\section{Boltzmann formalism}
\label{sec:Boltzmann}

To use the kinetic theory, one needs to assume that quasiparticles exist even in the LAB non-Fermi liquid, which, of course, is not true. Nevertheless, the $\epsilon$-expansion is precisely devised so that this type of problem is evaded: assuming the interaction between particles to be $\sim \epsilon$ and small, the diagrams for any desired quantity are to be computed using the non-interacting propagators. At a small value of the parameter $\epsilon$, the lengthscale where the quasiparticle description ceases to be valid behaves as $ \sim e^{1/\epsilon}$, and is then set finite only at the end of the calculation. Keeping this lengthscale as the longest in the problem, at all intermediate scales one has well-defined quasiparticles. Within the context of the hydrodynamic transport, this standard strategy from the theory of critical phenomena was first applied to the problem of universal conductivity at the superfluid-insulator transition \cite{Damle1997}.

The quasiparticles in the $\lambda$- energy band are described by the distribution function $f_{\lambda}$, where $f_{-,k}$ ($f_{+,k}$) is the distribution function of holes (electrons). The distribution functions of the quasiparticles determine the transport properties of the system, since the intraband-contribution of the electrical conductivity $j_{\alpha}$, and of the energy-stress tensor $\tau_{\alpha \beta}$ are given by
\begin{eqnarray}
 j_{\alpha}
 &=&
 e\sum_{\lambda} \int \frac{d^dk}{(2\pi)^d} \: v^{\alpha}_{\boldsymbol{k},\lambda} f_{\lambda}\\
 \tau_{\alpha \beta }
 &=&
 \sum_\lambda \int \frac{d^dk}{(2 \pi)^d}\:  v^{\alpha}_{\boldsymbol{k},\lambda} k^{\beta} f_{\lambda}
 \:.
\end{eqnarray}
These intraband contributions to the charge and momentum current are the ones that dominate the hydrodynamic transport regime and the interband processes (which correspond in the case of the electrical conductivity to the so-called \textit{Zitterbewegung}) to the currents are neglected.
The distribution function of the quasiparticles can be evaluated using the quantum Boltzmann equation. The Boltzmann equation is given by
\begin{equation}
 \frac{\partial f_{\lambda}}{\partial t}+ \mathbf{v}_{k,\lambda} \cdot \frac{\partial f_{\lambda}}{\partial \mathbf{x}} + \mathbf{F} \cdot \frac{\partial f_{\lambda}}{\partial \mathbf{k}}
 =
 I^{ee}[f_{\lambda}]\:,
\end{equation}
where $\boldsymbol{F}=e (\boldsymbol{E}+\boldsymbol{v}_{\lambda,k} \times \boldsymbol{B})$ is the Lorentz force, and $v^{\alpha}_{\lambda,k}=\partial \epsilon_{\mathbf{k},\lambda}/\partial k_{\alpha}=2 \lambda k_{\alpha}/(2m)$ is the velocity of the quasiparticles. $I^{ee}[f_{\lambda}]$ is the collision operator acting on the distribution function of the quasiparticles. It describes the changes of the distribution function due to scattering processes induced by an external perturbation and sets the inverse scattering time of the conductivity, $\sigma=\frac{e^2}{m} n \tau_{\sigma}$, and viscosity, $\eta \propto n_{\mathcal{E}} \tau_{\eta}$, where $n_{\mathcal{E}}$ is the energy density of the system.
\subsection{Collision integral}
\label{subsec:collisionint}
The collision integral is derived using the Keldysh-formalism, which relates the product of the greater (lesser) Green's function with the lesser (greater) self-energy to the collision integral \cite{Danielewicz1984}, i.e.,
$
 {I^{ee}(k,\omega)= \Sigma^>(k,\omega) G^<(k,\omega)-\Sigma^<(k,\omega) G^>(k,\omega)}
$. The self-energy is defined by the Born diagrams consisting of the direct and the exchange self-energy. (Further details can be found in Appendix~\ref{sec:quantumkineticeq}.) This derivation leads to the following expression for the collision integral:
\begin{widetext}
\begin{align}
I^{ee}[f_{\mu}(\epsilon_{\boldsymbol{k}})]
= &
(2\pi)^{d+1-3d}\int d^{d}k_{1}d^{d}k_{2}d^{d}k_{3}\,\delta(\mu\epsilon_{\boldsymbol{k}}+\lambda\epsilon_{\boldsymbol{k}_{\boldsymbol{1}}}-\rho\epsilon_{\boldsymbol{k_2}}-\nu\epsilon_{\boldsymbol{k_3}})\delta(\boldsymbol{k}+\boldsymbol{k}_{1}-\boldsymbol{k}_{2}-\boldsymbol{k}_{3}) \nonumber \\
\times & \{V(\boldsymbol{k-k_{2}})^{2} R_{\mu \rho \lambda \nu} (\boldsymbol{k},\boldsymbol{k_1},\boldsymbol{k_2}, \boldsymbol{k_3})
-V(\boldsymbol{k-k_{2}})V(\boldsymbol{k-k_{3}})Q_{\mu\rho \lambda \nu}(\boldsymbol{k},\boldsymbol{k_1}, \boldsymbol{k_{2}},\boldsymbol{k_3})\}\\
\times &
\{
\left[1-f_{\mu}(\epsilon_{\boldsymbol{k}})\right]\left[1-f_{\lambda}(\epsilon_{\boldsymbol{k_{1}}})\right]f_{\rho}(\epsilon_{\boldsymbol{k_{2}}})f_{\nu}(\epsilon_{\boldsymbol{k_{3}}})
-f_{\mu}(\epsilon_{\boldsymbol{k}})f_{\lambda}(\epsilon_{\boldsymbol{k_{1}}})[1-f_{\rho}(\epsilon_{\boldsymbol{k_{2}}})][1-f_{\nu}(\epsilon_{\boldsymbol{k_{3}}})]
 \}
 \:,  \nonumber
\end{align}
\begin{figure}
 \includegraphics[width=\textwidth]{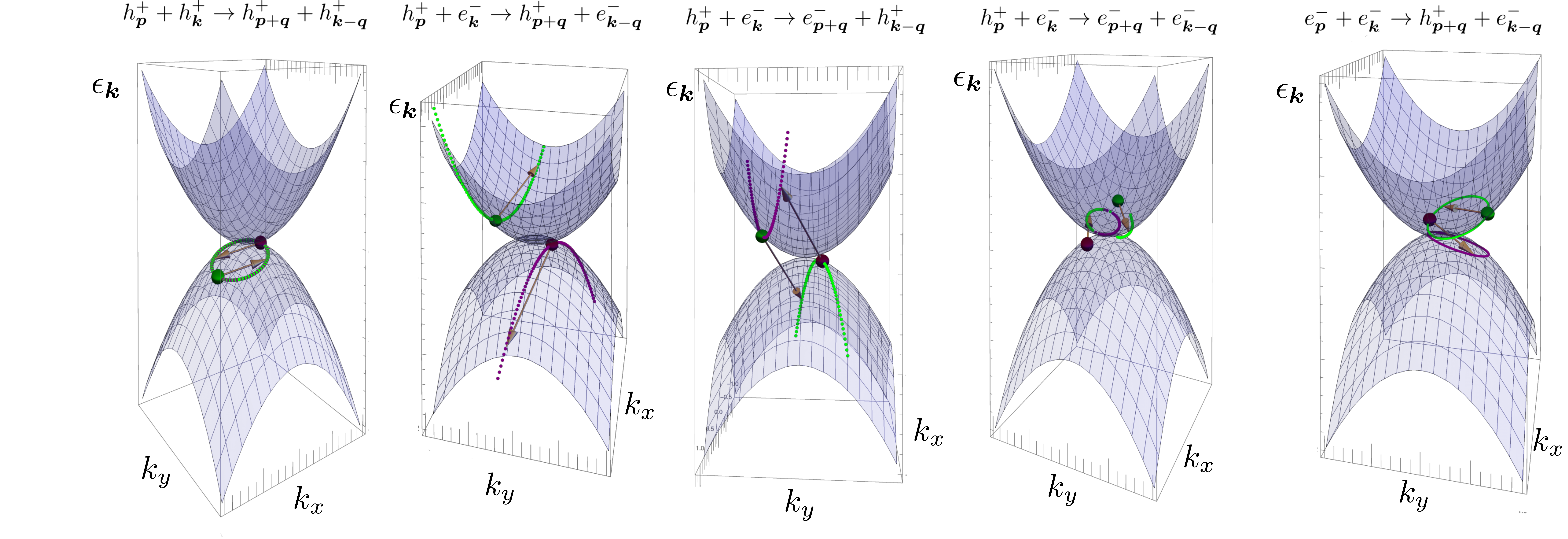}
 \caption{Allowed scattering processes in Luttinger semimetals. The energy dispersion dependent on the momentum component $k_x, k_y$ is plotted for fixed $\{p_z,k_z,q_z\}=0$. The green and purple dots indicate the initial quasiparticles, while the green and purple lines indicate all possible final scattering states. The first scattering process from the left describes the intraband scattering of two holes (two electrons), while the second and third from the left describe interband scattering, where the particle-hole number stays conserved. The last two scattering processes are collisions where the particle and hole number gets violated. They contribute significantly to the collision integral.}
 \label{fig:Fig1}
\end{figure}
\end{widetext}
which describes the scattering of two quasiparticles in the initial state $\boldsymbol{k}$ and $\boldsymbol{k_1}$ into the final state $\boldsymbol{k_2}$ and $\boldsymbol{k_3}$ and vice versa. The scattering amplitude of this process is defined by the Born diagrams for the self-energy, where $V(\boldsymbol{q})=4\pi e^2/|\boldsymbol{q}|^2 $. The functions $R_{\mu \rho \lambda \nu}(\boldsymbol{k},\boldsymbol{k_1}, \boldsymbol{k_2}, \boldsymbol{k_3})$ and $Q_{\mu \rho \lambda \nu}(\boldsymbol{k}, \boldsymbol{k_1}, \boldsymbol{k_2}, \boldsymbol{k_3})$ project the Coulomb interaction onto the eigenstates of the system and consist of the trace over a product of $16\times16$ matrices in $d=4$-dimensions. This expression of the scattering amplitude differs significantly from the scattering amplitude used by Dumitrescu  in Ref.~\onlinecite{PhysRevB.92.121102}, where the squared Coulomb potential, $|V(\mathbf{q})|^2$, was assumed for the scattering amplitude, and the projection onto the eigenstates of the system was not taken into account. This is the main difference between our two calculations, and also the main reason for the different numerical coefficient we eventually found.

The collision integral describes five different scattering processes. The first three intraband and interband scattering processes, depicted in Fig.~\ref{fig:Fig1}, conserve the particle and hole number. Two holes (electrons) scatter into the state of two holes (electrons) with different momenta or an electron and hole pair exchange the momentum $\mathbf{q}$ upon the collision process. These scattering channels correspond to the three scattering channels of graphene, where the linear energy spectrum in combination with the energy conservation only permits these processes.
However, in the Luttinger semimetals due to the quadratic energy dispersion two additional scattering channels, the so-called Auger processes, are allowed.
These are the processes where a particle and a hole scatter and the final state are two holes or two electrons and vice versa. These processes have a large  numerical contribution to the collision integral, which can be related to the inverse scattering time. Hence these additional scattering channels reduce the numerical coefficient of the corresponding transport properties.
\subsection{Linearization}
\label{subsec:Linearization}
In order to determine the distribution function of the quasiparticles, the quantum Boltzmann equation is linearized by choosing the following \textit{Ansatz}, i.e. $f_{\lambda}=f^{(0)}_{\lambda} + \delta f_{\lambda}$. Thereby, the distribution function is split into a part describing the equilibrium state of the system given by the Fermi-Dirac function, $f^{(0)}_{\lambda}=1/[\exp(\beta \epsilon_{\boldsymbol{k},\lambda}) +1]$  with $\beta=1/(k_B T)$, and an out-of-equilibrium correction
$\delta f_{\lambda}= \beta f^{(0)}_{\lambda}(1-f^{(0)}_{\lambda}) h(\boldsymbol{k},\lambda)$, where the function $h(\boldsymbol{k},\lambda)$ describes the coupling of the system to an external perturbation inducing the transport of the quasiparticles. In the case of the electrical conductivity $\sigma$ and of the shear viscosity $\eta$, we choose the respective \textit{Ansatz} \cite{Link2018a}
\begin{eqnarray}
 h_{\sigma}(\boldsymbol{k},\lambda)
 &=&
 v^{\alpha}_{\lambda,\boldsymbol{k}} E^{\alpha} g(k)
 \:,\label{eq:eq11}\\
 h_{\eta}(\boldsymbol{k},\lambda)
 &=&
 \lambda I_{\alpha \beta} X_{\alpha \beta} g(k)
 \:,\label{eq:eq12}
\end{eqnarray}
where the external perturbations are the electrical field $E^{\alpha}$ coupling to the velocity component $v^{\alpha}_{\lambda,k}$, and the external velocity gradient $X_{\alpha \beta}=(\frac{\partial u_{\alpha}}{\partial x_{\beta}} + \frac{\partial u_{\beta}}{\partial x_{\alpha}} - \frac{2}{d} \delta_{\alpha \beta} \partial_{\alpha} u_{\alpha})$ coupling to $I_{\alpha \beta}=(k_{\alpha} k_{\beta}-\frac{1}{d} k^2 \delta_{\alpha \beta})$. The function $g(k)$ is a superposition of the Laguerre-polynomials, i.e. ${g(k)=\sum_{n} \psi_{n} L^{(3)}_{\mu}(k^2)}$. Upon using this ansatz of the distribution function for the linearization, the Boltzmann equation can be cast into the matrix form, ${| L \rangle = I^{ee} | h \rangle}$, where the left-hand side denotes the Liouville operator acting on the distribution and the right-hand side describes the collision operator acting on $h$. Using the Hermiticity of the collision integral with respect to its inner product, the matrix equation can be recast to the following optimization problem, ${Q[h]=\langle h,L \rangle- \langle h|I^{ee}| h\rangle/2}$ where the expression is maximized with respect to $h$.
\section{Results}
\label{sec:Results}
\begin{figure}
 \includegraphics[width=\columnwidth]{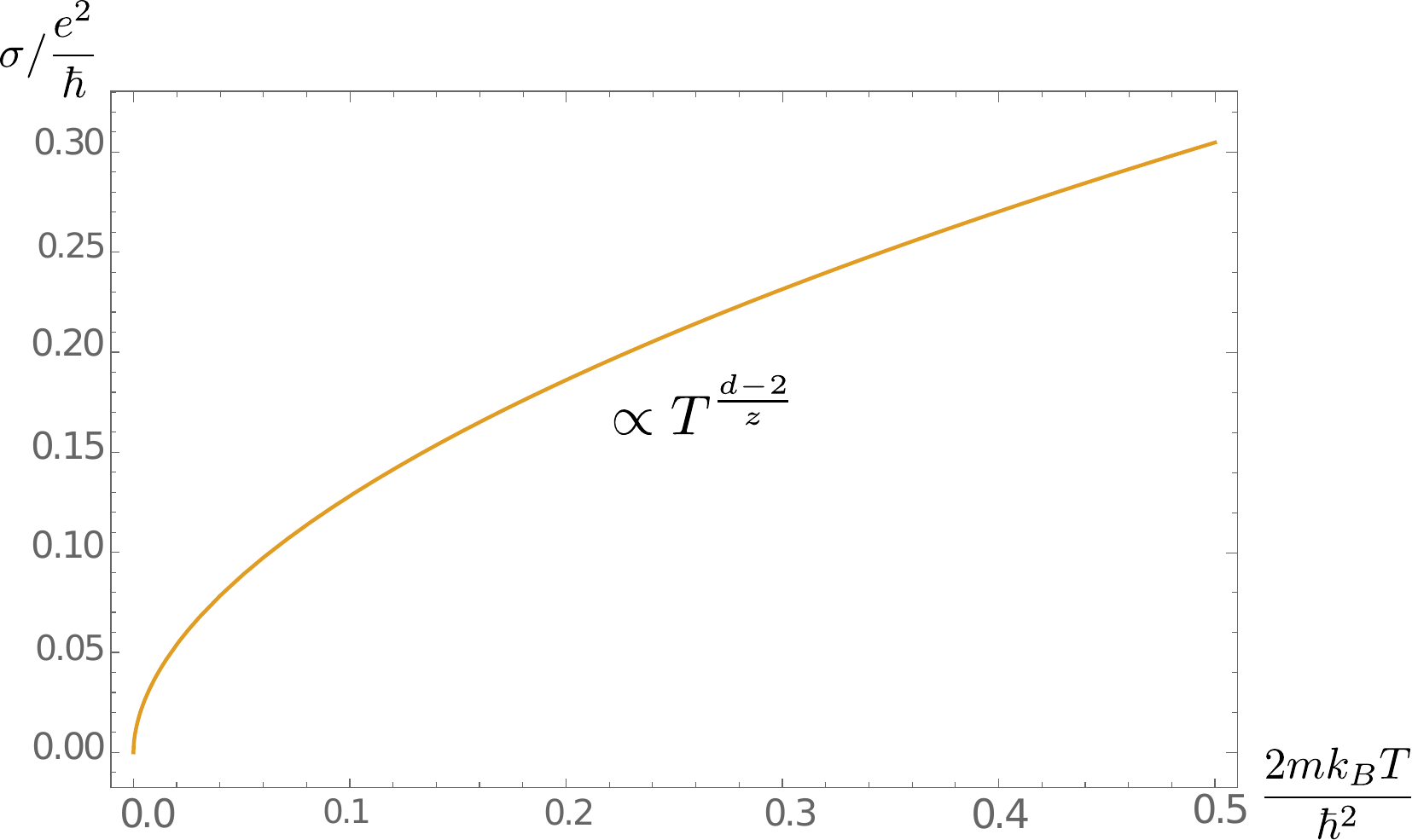}
 \caption{Temperature dependence of the dc electrical conductivity in $d=3$ dimensions with $\mathcal{C}_{\sigma}(\epsilon=0)$. The electrical conductivity decreases as a power-law with decreasing temperature.}
 \label{fig:Fig2}
\end{figure}

After evaluating the Boltzmann equation in $d=4$ and determining the form of the distribution function, the electrical conductivity is to the leading order in $\epsilon$  given by
\begin{equation}
\sigma_{\alpha \alpha}(T)
=
\frac{e^2}{\hbar}
\mathcal{C}_{\sigma}(\epsilon =0)
\left(\frac{2 m k_B T}{\hbar^2} \right)^{(d-2)/z} \frac{1}{e_*^4}
\end{equation}
with the numerical coefficient
\begin{equation}
 \mathcal{C}_{\sigma}(0) =12.840
 \:.
\end{equation}
Here $ e^2_* = 2 \pi {\tilde{e}}^2_{*} = 12 \pi \epsilon /( 3 N_d +1 )$.
The interacting Luttinger semimetal shows the power-law temperature dependence of the conductivity, as shown in Fig.~\ref{fig:Fig2}.  This should be contrasted with graphene where the Coulomb interaction leads to a logarithmic divergence of the dc conductivity at low temperatures\cite{Fritz2008}.

The shear viscosity to the leading order in interaction has the form
\begin{equation}
\eta(T)
=
\hbar\:
\mathcal{C}_{\eta}(\epsilon=0) \:
\left(\frac{2m k_B T}{\hbar^2} \right)^{d/z} \frac{1}{e_*^4}
\end{equation}
with the numerical coefficient
\begin{equation}
\mathcal{C}_{\eta}(0)=0.086
\:.
\end{equation}
This numerical coefficient is two orders of magnitude smaller than the value claimed earlier in Ref.~\onlinecite{PhysRevB.92.121102}, where the corresponding value of  $3.1$ was found. The main reason for this discrepancy is the different expression for the scattering amplitude, which is here not the bare Coulomb interaction, but the Coulomb interaction projected on the eigenstates of the system, while taking the direct and exchange self-energy into account. The entropy of the Luttinger semimetals has the same temperature dependence as the viscosity with
\begin{equation}
s(T)=k_B \:\mathcal{C}_{s}(\epsilon) \left(\frac{2 m k_B T}{\hbar^2}\right)^{d/z}
\end{equation}
and the numerical value $\mathcal{C}_{s}(0)=9 N_d \zeta(3)/(32 \pi^2)$ in $d=4$. Upon inserting the value of the fixed point defined in Eq.~\eqref{eq:fixedpoint}, the ratio of viscosity over entropy yields
\begin{equation}
\frac{\eta}{s}=\frac{\hbar}{4 \pi k_B} \frac{1.734}{\epsilon^2}
\:,
\end{equation}
which is the result announced in Eq.~\eqref{eq:result_visc}.
The ratio is diverging for $d=4$ dimension, since the system is at a non-interacting fixed point, which leads to a diverging viscosity. Setting $\epsilon=1$, which corresponds to $d=3$, the ratio comes close to the lower bound which indicates that the LAB non-Fermi liquid is close to being a ``nearly perfect fluid". However, this is the result obtained in the first-order calculation. Higher orders in $\epsilon$ might change the result significantly, as well as in a yet unknown direction. To go beyond leading order, the three-particle scattering processes need to be taken into account in the collision integral. Compounding the difficulty, the loop integrals need to be computed using the non-relativistic propagators of the Luttinger semimetal. This is considerably more difficult than in relativistic problems, such as Gross-Neveu-Yukawa theories at the critical point \cite{Mihaila2017, Ihrig2019}, for example.  We are not aware of any such higher-order calculations  for the hydrodynamic transport even in the simplest $\phi^4$-theory, which describes the superfluid-insulator transition\cite{Damle1997}. Also, the next-order term in the beta-function for the charge coupling (Eq.~\ref{eq:floweq}) in the Luttinger system is not known at the moment.
\section{Summary}
\label{sec:Summary}
We have combined the expansion around the upper critical dimension together with the Boltzmann equation to address the shear viscosity and the electrical conductivity at the Luttinger-Abrikosov-Beneslavskii (LAB) non-Fermi liquid fixed point of the interacting electronic system with the chemical potential at the quadratic band touching point in the energy spectrum. We found the projection of the Coulomb interaction onto the eigenstates of the system to be important and to have a significant quantitative  effect on the numerical values of the transport coefficients. Additional scattering processes, the so-called Auger processes, which violate conservation of the particle and hole number, are also allowed, and have been fully included. The interacting Luttinger semimetal shows a power-law temperature dependence of the dc electrical conductivity that reflects the non-trivial value of the dynamical exponent, and may, for example,  be used to extract it experimentally. The ratio of viscosity over entropy was computed near four dimensions, and it was found that when the leading-order result is continued to $d=3$, it brings it close to, but nevertheless above, the Kovtun-Son-Starinets lower bound. The LAB non-Fermi liquid by this measure should therefore be considered as a strongly interacting ``nearly perfect fluid".

Some interesting open questions for future studies are the influence of impurities, which are relevant at the LAB fixed point and will modify the relaxation rate of the scattering processes, as well as the influence of a magnetic field on the transport quantities. The corresponding Hall conductivity and Hall viscosity could have a significant influence on the flow of the quasiparticles and might lead to effects such as a strong positive magneto resistance.
Also it will be interesting to see if other computational strategies such as the Mori-Zwanzig memory formalism \cite{Zaanen2015,Hartnoll2018}, or the Kubo formalism \cite{Bradlyn2012} are able to go beyond our obtained first order result.

\section{Acknowledgment}

We thank J. Schmalian and E. I. Kiselev for fruitful discussions. J. M.L. was supported by the DFG grant No.  LI 3628/1-1. I.F.H. was supported by the NSERC of Canada.

\appendix

\section{The system}
\label{ap:theSystem}
To determine the transport properties of the Luttinger semimetals, we study the Hamiltonian of the form
\begin{eqnarray}
 H
 &=&
 H_0+H_{int} \nonumber\\
 &=&
 \psi^{\dagger}\left(\frac{1}{2m}\sum_{a=1}^{f_d} d_a(\boldsymbol{p}) \gamma_a\right) \psi  \label{eq:Hamiltonian_used} \\
 &+&
 \frac{1}{2} \int d^d r d^d r'
 \psi^{\dagger}_{\boldsymbol{r}'} \psi^{\dagger}_{\boldsymbol{r}}
 \frac{e^2}{|\boldsymbol{r}-\boldsymbol{r}'|^{d-2}} \psi_{\boldsymbol{r}}\psi_{\boldsymbol{r}'}
 \:, \nonumber
\end{eqnarray}
where $H_0$ describes a non-interacting system with a quadratic energy band touching point and the $H_{int}$ the Coulomb interaction between the quasiparticles. The choice of this maximally symmetric Hamiltonian can be justified, since the additional terms in the non-interacting part of the Hamiltonian $H_0$ which introduce a particle-hole asymmetry and the reduction of the rotational down to cubic symmetry are irrelevant at the LAB fixed point \cite{Boettcher2017}. In the following we set $2 m=1$ and insert it appropriately at the end of the calculation. The functions $d_{a}(\boldsymbol{p})$ are the spherical harmonics for the angular momentum of 2 which are defined as
$d_a(\boldsymbol{p})=\sqrt{\frac{d}{2(d-1)}} p_i \left( \Lambda^a \right)_{ij} p_j$, where $\Lambda^a$ are the real Gell-Mann matrices. In $d=3$, there are five independent spherical harmonics with $f_3=5$, while in $d=4$ the number increases to 9 with $f_4=9$ \cite{Janssen2015}. In the following, the explicit form of the $\gamma$-matrices in $d=3$ and $d=4$, and the expression of the spherical harmonics, are given.
The $\gamma$-matrices obeying the Clifford-algebra are defined in $d=3$ as
\begin{eqnarray}
 \gamma_1 &=& \sigma_1 \otimes \mathbb{1}_{2\times 2} \nonumber \\
 \gamma_2 &=& \sigma_3 \otimes \sigma_3 \nonumber \\
 \gamma_3 &=& \sigma_3 \otimes \sigma_1\\
 \gamma_4 &=& \sigma_3 \otimes \sigma_2 \nonumber \\
 \gamma_5 &=& \sigma_2 \otimes \mathbb{1}_{2 \times 2} \nonumber
 \:,
\end{eqnarray}
where $\sigma_i$ are the Pauli matrices. These $\gamma$-matrices can be extended to $d=4$ by implementing the following procedure: \\
First define
\begin{equation}
 \Gamma_a=
 \begin{cases}
  \sigma_3 \otimes \gamma_a\\
  \sigma_1 \otimes \mathbb{1}_{4\times 4}\\
  \sigma_2 \otimes \mathbb{1}_{4\times 4}\:,
  \end{cases}
\end{equation}
which lead to the $\gamma$-matrices in $d=4$
\begin{equation}
 G_a=
 \begin{cases}
  \sigma_3 \otimes \Gamma_a\\
  \sigma_1 \otimes \mathbb{1}_{8\times8}\\
  \sigma_2 \otimes \mathbb{1}_{8 \times 8}
  \end{cases}
  \:.
\end{equation}
The Gell-Mann matrices in $d=4$ are given by
\begin{widetext}
 \begin{eqnarray}
 \Lambda^1
 &=&
 \left(
\begin{array}{cccc}
 1 & 0 & 0 & 0 \\
 0 & -1 & 0 & 0 \\
 0 & 0 & 0 & 0 \\
 0 & 0 & 0 & 0 \\
\end{array}
\right)
\:, \quad
 \Lambda^2=\left(
\begin{array}{cccc}
 0 & 1 & 0 & 0 \\
 1 & 0 & 0 & 0 \\
 0 & 0 & 0 & 0 \\
 0 & 0 & 0 & 0 \\
\end{array}
\right)
\:, \quad
\Lambda^3= \left(
\begin{array}{cccc}
 0 & 0 & 1 & 0 \\
 0 & 0 & 0 & 0 \\
 1 & 0 & 0 & 0 \\
 0 & 0 & 0 & 0 \\
\end{array}
\right)
\:, \nonumber \\
 \Lambda^4
 &=&
 \left(
\begin{array}{cccc}
 0 & 0 & 0 & 0 \\
 0 & 0 & 1 & 0 \\
 0 & 1 & 0 & 0 \\
 0 & 0 & 0 & 0 \\
\end{array}
\right)
\:,\quad
 \Lambda^5=
 \frac{1}{\sqrt{3}}\left(
\begin{array}{cccc}
 -1 & 0 & 0 & 0 \\
 0 & -1 & 0 & 0 \\
 0 & 0 & 2 & 0 \\
 0 & 0 & 0 & 0 \\
\end{array}
\right)
\:,\quad
 \Lambda^6=\left(
\begin{array}{cccc}
 0 & 0 & 0 & 1 \\
 0 & 0 & 0 & 0 \\
 0 & 0 & 0 & 0 \\
 1 & 0 & 0 & 0 \\
\end{array}
\right)
\:, \nonumber \\
 \Lambda^7
 &=&
 \left(
\begin{array}{cccc}
 0 & 0 & 0 & 0 \\
 0 & 0 & 0 & 1 \\
 0 & 0 & 0 & 0 \\
 0 & 1 & 0 & 0 \\
\end{array}
\right)
\:, \quad
 \Lambda^8=\left(
\begin{array}{cccc}
 0 & 0 & 0 & 0 \\
 0 & 0 & 0 & 0 \\
 0 & 0 & 0 & 1 \\
 0 & 0 & 1 & 0 \\
\end{array}
\right)
\:, \quad
\Lambda^9= \frac{1}{\sqrt{6}}\left(
\begin{array}{cccc}
 -1 & 0 & 0 & 0 \\
 0 & -1 & 0 & 0 \\
 0 & 0 & -1 & 0 \\
 0 & 0 & 0 & 3 \\
\end{array}
\right)
\:.
\end{eqnarray}
\end{widetext}
In this paper all calculations, which are performed to determine the collision integral and thus the distribution function of the quasiparticles in the Boltzmann formalism, are done in $d=4$. The reason for this is that the upper critical dimension of the system is $d=4$ and that the Coulomb interaction is a  relevant interaction in $d=3$ leading to a quantum critical phase, the LAB phase \cite{Abrikosov1974,Moon2013}. To be able to perform a RG analysis, the small parameter $\epsilon=4-d$ is introduced. For small $\epsilon$, the existence of quasiparticles is a valid assumption.
Hence, the Hamiltonian in $d=4$ dimensions is given by a $16\times 16$ matrix and the valence and conductance bands have a degeneracy of $N_{d=4}=8$. The Hamiltonian $H_0$ defined in Eq.~\eqref{eq:Hamiltonian_used} is diagonalized by the transformation matrix $P$ with
\begin{equation}
 P^{-1}(\boldsymbol{k}) H_0(\boldsymbol{k}) P(\boldsymbol{k})
 =
 (\sigma_3 \otimes \mathbb{1}_{8\times 8}) \epsilon_{\boldsymbol{k}}
 \:,
\end{equation}
with $\epsilon_{\boldsymbol{k}}=k^2/(2m)$. The matrix $P(\boldsymbol{k})$ transforms the system into its eigenstate basis, i.e., the basis that describes the creation and annihilation of the quasiparticles in the corresponding energy band.
The Green's function $G$ and the self-energy $\Sigma$, defined over the Hamiltonian $H_0$, are also diagonalized by the same transformation matrix $P$, and we find
\begin{equation}
 P^{-1} G P =g \:, \qquad  P^{-1} \Sigma P =\sigma
 \:.
\end{equation}
\section{The quantum kinetic equation}
\label{sec:quantumkineticeq}
\begin{figure}[t]
 \includegraphics[width=\columnwidth]{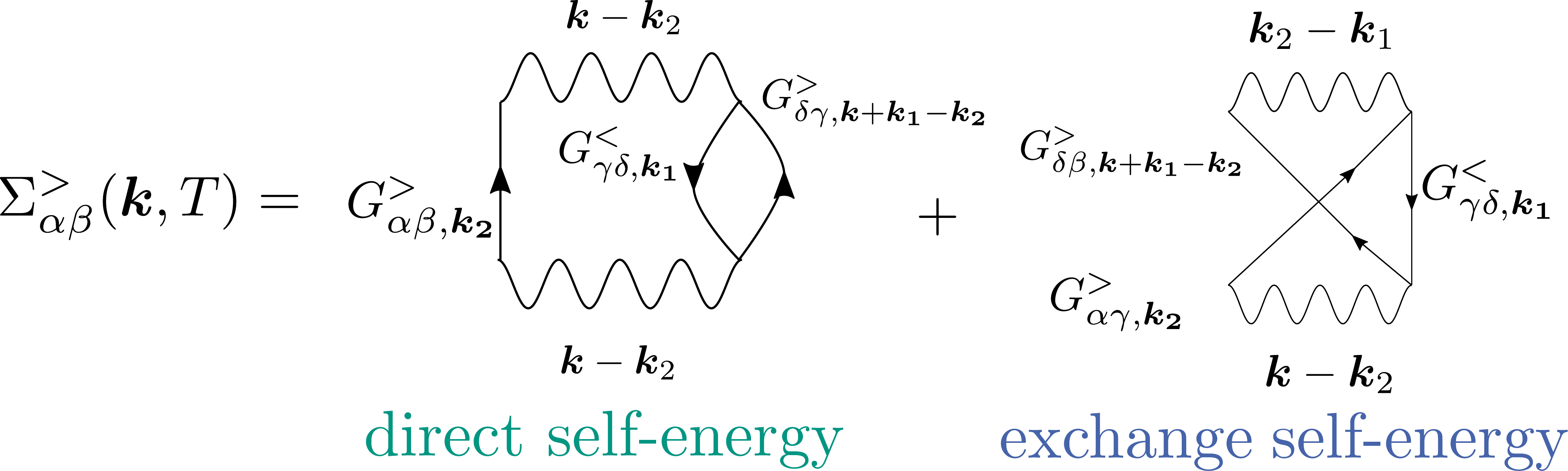}
 \caption{Born diagrams for the greater self-energy. The lesser self-energy is given by similar diagrams only the lesser and greater Green's functions are exchanged.}
 \label{fig:Borndiagram}
\end{figure}
Next, the Keldysh formalism is used. The diagonal elements of the lesser and greater Green's function in the energy band basis can be related to the distribution functions of the quasiparticles by
\begin{eqnarray}
 g^{<}_{\lambda \lambda'}
 &=&
 i f_{\lambda}(\omega) A(\boldsymbol{p},\omega) \delta_{\lambda \lambda'}\\
  g^{>}_{\lambda \lambda'}
 &=&
 -i [1-f_{\lambda}(\omega)] A(\boldsymbol{p},\omega) \delta_{\lambda \lambda'}\:,
\end{eqnarray}
where $A(\boldsymbol{p},\omega)$ is the spectral function which we assume here for simplicity as $A(\boldsymbol{p},\omega)=2\pi \delta(\omega-\epsilon_{\boldsymbol{p},\lambda})$. The equation of motion for the lesser Green's function can be identified with the Boltzmann equation \cite{Danielewicz1984} and holds
\begin{equation}
  \int \frac{d \omega}{2 \pi} (\partial_t +\boldsymbol{v}_{\lambda,\boldsymbol{k}} \cdot \frac{\partial}{\partial \boldsymbol{x}}+\boldsymbol{F}\cdot \frac{\partial}{\partial \boldsymbol{k}})(i g^<)
 =
 \int_{\omega} (\sigma^> g^<-\sigma^< g^>)
 \:.
\end{equation}
To determine the expression of the collision integral, the expression for the lesser and greater self-energy must be found. These self-energies are defined by the Born diagrams shown in Fig.~\ref{fig:Borndiagram} and we find for the greater self-energy the expression
\begin{eqnarray}
& & \Sigma_{\alpha\beta}^{>}(k,\omega,T) \nonumber \\
& & \quad=
\int\frac{d^{d}k_{1}}{(2\pi)^{d}}\frac{d\omega_{1}}{2\pi}\frac{d^{d}k_{2}}{(2\pi)^{d}}\frac{d\omega_{2}}{2\pi}\frac{d^{d}k_{3}}{(2\pi)^{d}}\frac{d\omega_{3}}{2\pi}
 \\
& & \quad\times
 (2\pi)^{d+1}\delta(\boldsymbol{k}+\boldsymbol{k}_{1}-\boldsymbol{k}_{2}-\boldsymbol{k}_{3})\delta(\omega+\omega_{1}-\omega_{2}-\omega_{3})
  \nonumber \\
 &\times&
 \{V(\boldsymbol{k-k_{2}})^{2}G_{\gamma\delta}^{<}(\omega_{1},\boldsymbol{k}_{1})G_{\alpha\beta}^{>}(\omega_{2},\boldsymbol{k}_{2})G_{\delta\gamma}^{>}(\omega_{3},\boldsymbol{k}_{3})  \nonumber \\
 &-&
 V(\boldsymbol{k-k_{2}})V(\boldsymbol{k-k_{3}})G_{\gamma\delta}^{<}(\omega_{1},\boldsymbol{k}_{1})G_{\alpha\gamma}^{>}(\omega_{2},\boldsymbol{k}_{2})G_{\delta\beta}^{>}(\omega_{3},\boldsymbol{k}_{3})\}
 \:,  \nonumber
\end{eqnarray}
where $V(\boldsymbol{q})=4\pi e^2/|\mathbf{q}|^2$.
To obtain the expression for the lesser self-energy, the lesser and greater Green's functions need to be exchanged.
Hence, we obtain as expression for the collision integral the following form:
\begin{widetext}
\begin{align}
I^{ee}[f_{\mu}(\epsilon_{\boldsymbol{k}})]= & \int\frac{d\omega}{2\pi}\left(-i\sigma_{\mu\mu}^{<}[1-f_{\mu}(\epsilon_{\boldsymbol{k}})]-i\sigma_{\mu\mu}^{>}f_{\mu}(\epsilon_{\boldsymbol{k}})\right)(2\pi)\delta(\omega-\epsilon_{\boldsymbol{k},\mu})
\nonumber \\
= & (2\pi)^{d+1-3d}\int d^{d}k_{1}d^{d}k_{2}d^{d}k_{3}\,\delta(\mu\epsilon_{\boldsymbol{k}}+\lambda\epsilon_{\boldsymbol{k}_{\boldsymbol{1}}}-\rho\epsilon_{\boldsymbol{k_2}}-\nu\epsilon_{\boldsymbol{k_3}})\delta(\boldsymbol{k}+\boldsymbol{k}_{1}-\boldsymbol{k}_{2}-\boldsymbol{k}_{3})
 \\
\times &
\{V(\boldsymbol{k-k_{2}})^{2} R_{\mu \rho \lambda \nu}(\boldsymbol{k},\boldsymbol{k_1},\boldsymbol{k_2},\boldsymbol{k_3})
-V(\boldsymbol{k-k_{2}})V(\boldsymbol{k-k_{3}}) Q_{\mu \rho \lambda \nu}(\boldsymbol{k},\boldsymbol{k_1},\boldsymbol{k_2},\boldsymbol{k_3})  \}
 \nonumber \\
\times &
\{\left[1-f_{\mu}(\epsilon_{\boldsymbol{k}})\right]\left[1-f_{\lambda}(\epsilon_{\boldsymbol{k_{1}}})\right]f_{\rho}(\epsilon_{\boldsymbol{k_{2}}})f_{\nu}(\epsilon_{\boldsymbol{k_{3}}})
-
f_{\mu}(\epsilon_{\boldsymbol{k}})f_{\lambda}(\epsilon_{\boldsymbol{k_{1}}})[1-f_{\rho}(\epsilon_{\boldsymbol{k_{2}}})][1-f_{\nu}(\epsilon_{\boldsymbol{k_{3}}})]
\}
\:, \nonumber
\end{align}
\end{widetext}
where the functions $R_{\mu \rho \lambda \nu}$ and $Q_{\mu \rho \lambda \nu}$, which project the Coulomb interaction on the eigenstates of the system, are given by:
\begin{eqnarray}
& & R_{\mu \rho \lambda \nu}(\boldsymbol{k},\boldsymbol{k_1},\boldsymbol{k_2},\boldsymbol{k_3})\\
& & \quad =
 M_{\mu\rho}(\boldsymbol{k},\boldsymbol{k_{2}})M_{\lambda\nu}(\boldsymbol{k}_{1},\boldsymbol{k_{3}})M_{\rho\mu}(\boldsymbol{k_{2},k})M_{\nu\lambda}(\boldsymbol{k_{3},k_{1}})
 \nonumber \\
& & \quad =
T_{\mu\rho}(\boldsymbol{k},\boldsymbol{k_{2}}) T_{\lambda\nu}(\boldsymbol{k}_{1},\boldsymbol{k_{3}})
 \:,
 \nonumber \\
& & Q_{\mu \rho \lambda \nu}(\boldsymbol{k},\boldsymbol{k_1},\boldsymbol{k_2},\boldsymbol{k_3}) \\
& & \quad = M_{\mu\rho}(\boldsymbol{k,k_{2}})M_{\lambda\nu}(\boldsymbol{k_{1},k_{3}})M_{\rho\lambda}(\boldsymbol{k_{2},k_{1}})M_{\nu\mu}(\boldsymbol{k_{3},k})
\:. \nonumber
\end{eqnarray}
The matrix $M_{\mu \rho}$ is defined as
\begin{equation}
 M_{\mu \rho}(\boldsymbol{k},\boldsymbol{k_1})
 =
 P^{-1}_{\mu \alpha}(\boldsymbol{k}) P_{\alpha \rho}(\boldsymbol{k_1})
 \:.
\end{equation}
The expression of the direct self-energy can be simplified further by identifying the term:
\begin{eqnarray}
 T_{\alpha\beta}(\boldsymbol{k},\boldsymbol{p})
 &=&
 M_{\alpha\beta}(\boldsymbol{k},\boldsymbol{p})M_{\beta\alpha}(\boldsymbol{p,k})\nonumber \\
 &=&
 \frac{N_{d}}{2}\left(1+\frac{\alpha\beta}{\epsilon_{k}\epsilon_{p}}d_{a}(\boldsymbol{p})d_{a}(\boldsymbol{k})\right)\\
 &=&
 \frac{N_{d}}{2}\left(1+\frac{\alpha\beta}{\epsilon_{\boldsymbol{k}}\epsilon_{\boldsymbol{p}}}\frac{1}{d-1}[d(\boldsymbol{k}\cdot\boldsymbol{p})^{2}-k^{2}p^{2}]\right)\:.
 \nonumber
\end{eqnarray}
Here, an additional comment is appropriate in order to clarify the meaning of the Greek indices appearing in the collision integral. The Greek indices, appearing in the distribution function and the $\delta$ distribution  imposing the energy-conservation, have the value of $-1$ or $+1$  and indicate the valence or conductance band of the system. The fact that these energy bands are degenerate is reflected in the Greek indices arising in $R_{\mu \rho \lambda \nu}$ and $Q_{\mu \rho \lambda \nu}$. In these functions whenever the matrix element $M_{\alpha=+1 \beta=-1}$ occurs, what is meant is that all matrix elements of the $16\times 16$ matrix $M_{ij}$ where $i\in\{1,2,3,4,5,6,7,8\}$ and $j \in \{9,10,11,12,13,14,15,16\}$ are summed up. (The same reasoning applies also for all other combinations of $\alpha$ and $\beta$.)
\section{Linearization of the Boltzmann equation}
In this appendix, we determine the distribution function of the quasiparticles when either an external electrical field $\boldsymbol{E}$ or an external velocity gradient is applied, $X_{\alpha \beta}=(\frac{\partial u_{\alpha}}{\partial x_{\beta}} + \frac{\partial u_{\beta}}{\partial x_{\alpha}} -\frac{2}{d} \delta_{\alpha \beta} \partial_{\alpha} u_{\alpha})$.
As first step, the Boltzmann equation is linearized by choosing the following \textit{Ansatz} for the distribution function:
\begin{equation}
 f_{\lambda}=f^{(0)}_{\lambda} + \delta f_{\lambda}
 \:,
\end{equation}
where $f^{(0)}_{\lambda}=1/(1+\exp(\beta \tilde{\epsilon}_{\boldsymbol{k},\lambda }))$ with $\tilde{\epsilon}_{\boldsymbol{k},\lambda }=\lambda \epsilon_{\boldsymbol{k}}$ for the conductivity and $\tilde{\epsilon}_{\boldsymbol{k},\lambda }=\lambda \epsilon_{\boldsymbol{k}}-\boldsymbol{k}\cdot \boldsymbol{u}(\boldsymbol{x})$ for the viscosity. The function $\delta f_{\lambda}$ describes the out-of-equilibrium correction to the distribution function and is defined as:
\begin{equation}
 \delta f_{\lambda}= \beta f_{\lambda}^{(0)} (1-f_{\lambda}^{(0)}) h(\boldsymbol{k},\lambda)
 \:,
\end{equation}
where $h(\boldsymbol{k},\lambda)$ describes the coupling of the system to the external perturbation. The exact expression of $h$ for the viscosity and the electrical conductivity is defined in Eq.~\eqref{eq:eq11} and \eqref{eq:eq12}. The function $g(k)$ appearing in both \textit{Ans\"atze} are defined as $g(k)=\sum_n \psi_n L_n^{(3)}(k^2)$, where the functions $L_n^{(3)}(k^3)$ are the associated Laguerre polynomials and $\psi_n$ are coefficients weighting the different polynomials.  After the linearization, the collision integral can be cast into the form:
\begin{widetext}
\begin{align}
 I^{lin}_{\mu}(\boldsymbol{k})&=\beta(2\pi)^{d+1-3d}\int d^{d}k_{1}d^{d}q\,\delta(\mu\epsilon_{\boldsymbol{k}}+\lambda\epsilon_{\boldsymbol{k}_{\boldsymbol{1}}}-\rho\epsilon_{\boldsymbol{k-q}}-\nu\epsilon_{\boldsymbol{k_{1}+q}}) \nonumber \\
 \times&\{V(\boldsymbol{q})^2 R_{\mu \rho \lambda \nu}(\boldsymbol{k},\boldsymbol{k_1},\boldsymbol{k}-\boldsymbol{q},\boldsymbol{k_1}+\boldsymbol{q})
-V(\boldsymbol{q})V(\boldsymbol{k-q-k_{1}})  Q_{\mu \rho \lambda \nu}(\boldsymbol{k},\boldsymbol{k_1},\boldsymbol{k}-\boldsymbol{q},\boldsymbol{k_1}+\boldsymbol{q})
 \} \\
 \times&\frac{h(\lambda,\boldsymbol{k_{1}})+h(\mu,\boldsymbol{k})-h(\nu,\boldsymbol{k_{1}}+\boldsymbol{q})-h(\rho,\boldsymbol{k}-\boldsymbol{q})}{(1+e^{\nu \beta \epsilon_{\boldsymbol{k_{1}+q}}})(1+e^{\rho \beta \epsilon_{\boldsymbol{k-q}}})(1+e^{-\mu \beta\epsilon_{\boldsymbol{k}}})(1+e^{-\lambda \beta \epsilon_{\boldsymbol{k}_{\boldsymbol{1}}}})}
 \:.
 \nonumber
\end{align}
\end{widetext}
\subsection{Viscosity}
We now focus on the evaluation of the shear viscosity of the system. We assume that the electrical field $\boldsymbol{E}$ and the magnetic field $\boldsymbol{B}$ is zero, that the distribution function is independent of the time $t$, and that the out-of equilibrium correction to the distribution function is given by ${\delta f_{\lambda}=\beta f^{(0)}_{\lambda} (1-f^{(0)}_{\lambda}) h_{\eta}(k,\lambda)}$ with ${h_{\eta}(k,\lambda)= \lambda I_{\alpha \beta}(k) g(k) X_{\alpha \beta}}$. The Boltzmann equation can be cast into the form:
\begin{equation}
 \lambda \hbar \beta 2 I_{\alpha \beta}(k)
 f^{(0)}_{\lambda} (1-f^{(0)}_{\lambda}) X_{\alpha \beta}
 =I_{\mu}^{lin,\eta}(\boldsymbol{k})
 \:.
\end{equation}
We modify the Boltzmann equation further by multiplying both sides by the term $I_{\alpha \beta}(\boldsymbol{k})$ and by using the following symmetry properties of the collision integral
\begin{equation}
 I^{ee}[f_{\lambda}(\epsilon_{\boldsymbol{k}})]=\int_P I_{\alpha \beta}(\mathbf{p}) Z(\mathbf{p,k})
 \:,
\end{equation}
which yields the following form of the Boltzmann equation
\begin{widetext}
\begin{eqnarray}
 \frac{ \hbar \beta 2  \: \lambda Y_{\boldsymbol{k},\boldsymbol{k}}}{[1+\exp{(\beta \epsilon_{\boldsymbol{k}})}][ 1+\exp{(-\beta\epsilon_{\boldsymbol{k}})}]}
 &=&
 \beta(2\pi)^{d+1-3d}\int d^{d}k_{1}d^{d}q\,\delta(\mu\epsilon_{\boldsymbol{k}}+\lambda\epsilon_{\boldsymbol{k}_{\boldsymbol{1}}}-\rho\epsilon_{\boldsymbol{k-q}}-\nu\epsilon_{\boldsymbol{k_{1}+q}}) \nonumber \\
 &\times&
 \{V(\boldsymbol{q})^2 R_{\mu \rho \lambda \nu}(\boldsymbol{k},\boldsymbol{k_1},\boldsymbol{k}-\boldsymbol{q},\boldsymbol{k_1}+\boldsymbol{q})
-V(\boldsymbol{q})V(\boldsymbol{k-q-k_{1}})  Q_{\mu \rho \lambda \nu}(\boldsymbol{k},\boldsymbol{k_1},\boldsymbol{k}-\boldsymbol{q},\boldsymbol{k_1}+\boldsymbol{q})
 \} \nonumber \\
 &\times&
 \frac{\lambda Y_{\boldsymbol{k},\boldsymbol{k_1}} g(k_1)+ \mu Y_{\boldsymbol{k},\boldsymbol{k}} g(k) - \nu Y_{\boldsymbol{k},\boldsymbol{k_1}+\boldsymbol{q}} g(k_1+q) - \rho Y_{\boldsymbol{k},\boldsymbol{k}-\boldsymbol{q}} g(k-q)}{(1+e^{\beta \nu \epsilon_{\boldsymbol{k_{1}+q}}})(1+e^{\beta \rho\epsilon_{\boldsymbol{k-q}}})(1+e^{-\beta \mu\epsilon_{\boldsymbol{k}}})(1+e^{-\beta \lambda\epsilon_{\boldsymbol{k}_{\boldsymbol{1}}}})}
 \:,
 \label{eq:Boltz_visc}
 \end{eqnarray}
with
\begin{equation}
 Y_{\boldsymbol{p},\boldsymbol{k}}
 =
 I_{\alpha \beta}(\boldsymbol{k}) I_{\alpha \beta}(\boldsymbol{p})
 =
 (\boldsymbol{p}\cdot \boldsymbol{k})^2-\frac{1}{d} p^2 k^2
 \:.
\end{equation}
The collision operator $I_{coll}[\tilde{h}_{\eta}]$ in Eq.~\eqref{eq:Boltz_visc} with $\tilde{h}_{\eta}=\lambda Y_{\boldsymbol{k},\boldsymbol{k}} g(k)$
is self-adjoint with respect to the scalar product
\begin{equation}
 \langle \tilde{h}_{\eta,1}| \tilde{h}_{\eta,2} \rangle
 =
 \sum_{\lambda} \int \frac{d^d k}{(2 \pi)^d} \tilde{h}_{\eta,1}(\boldsymbol{k}) \tilde{h}_{\eta,2}(\boldsymbol{k})
 \:.
\end{equation}
Using these properties, the above Boltzmann equation can be related to a functional $Q[g]$, so that solving Eq.~\eqref{eq:Boltz_visc} can be identified with finding the stationary point of
\begin{equation}
 \frac{\partial Q[g]}{\partial g}=0
 \:.
\end{equation}
The shear viscosity is given by the following functional $Q[g]$:
\begin{eqnarray}
 Q_{\eta}[g]
 &=&
 -\beta \frac{(2\pi)^{1-3d}}{2}\int d^{d}kd^{d}k_{1}d^{d}q\,\{\frac{\delta(\epsilon_{\boldsymbol{k}}-\epsilon_{\boldsymbol{k}_{\boldsymbol{1}}+\boldsymbol{q}}-\epsilon_{\boldsymbol{k-q}}+\epsilon_{\boldsymbol{k_{1}}})K^{(conserving)}(\boldsymbol{k,k_{1,}q})}{(1+e^{\beta \epsilon_{\boldsymbol{k_{1}+q}}})(1+e^{\beta\epsilon_{\boldsymbol{k-q}}})(1+e^{-\beta \epsilon_{\boldsymbol{k}}})(1+e^{-\beta \epsilon_{k_{\boldsymbol{1}}}})}
 \nonumber \\
&& \qquad \quad \quad \times
\left[Y_{\boldsymbol{k,k1}}g(k_{1})+Y_{\boldsymbol{k,k}}g(k)-Y_{\boldsymbol{k,k_{1}+q}}g(k_{1}+q)-Y_{\boldsymbol{k,k-q}}g(k-q)\right]^{2}
\nonumber \\
&&
\qquad \quad +
\frac{\delta(\epsilon_{\boldsymbol{k}}-\epsilon_{\boldsymbol{k}_{1}}-\epsilon_{\boldsymbol{k-q}}-\epsilon_{\boldsymbol{k_{1}+q}})  }{(1+e^{\beta \epsilon_{\boldsymbol{k_{1}+q}}})(1+e^{\beta \epsilon_{\boldsymbol{k-q}}})(1+e^{-\beta \epsilon_{\boldsymbol{k}}})(1+e^{\beta \epsilon_{\boldsymbol{k}_{1}}})}\\
&& \qquad \quad \quad
\times \big(
K_{ph\to pp}(\boldsymbol{k},\boldsymbol{k_1},\boldsymbol{q})
\left[ -Y_{\boldsymbol{k},\boldsymbol{k}_{1}}g(k_{1})+Y_{\boldsymbol{k,k}}g(k)-Y_{\boldsymbol{k},\boldsymbol{k_{1}+q}}g(k_{1}+q)-Y_{\boldsymbol{k},\boldsymbol{k-q}}g(k-q) \right]^2
\nonumber \\
&& \qquad \qquad \quad
+K_{hp\to pp}(\mathbf{k},\mathbf{k_1},\mathbf{q})
[+Y_{\boldsymbol{k},\boldsymbol{k}_{1}}g(k)-Y_{\boldsymbol{k_{1},k_{1}}}g(k_{1})-Y_{\boldsymbol{-k_{1}},-\boldsymbol{k+q}}g(k-q)-Y_{-\boldsymbol{k}_1,-\boldsymbol{k_1-q}}g(k_{1}+q)]^2
\nonumber \\
&&
\qquad \qquad \quad
- K_{pp\to ph}(\mathbf{k},\mathbf{k_1},\mathbf{q})
[-Y_{-\boldsymbol{k+q},\boldsymbol{k}_{1}}g(k_{1})-Y_{\boldsymbol{k-q,k-q}}g(k-q)
-Y_{-\boldsymbol{k+q},\boldsymbol{k_{1}+q}}g(k_{1}+q)+Y_{-\boldsymbol{k+q},-\boldsymbol{k}}g(k)]^2
\nonumber\\
&&
\qquad \qquad \quad
-K_{pp\to hp}(\mathbf{k},\mathbf{k_1},\mathbf{q})
[-Y_{\boldsymbol{-k_{1}},\boldsymbol{k}-\boldsymbol{q}}g(k-q)-Y_{\boldsymbol{k_{1},k_{1}}}g(k_{1})+Y_{-\boldsymbol{k_{1}},\boldsymbol{k}}g(k)-Y_{\boldsymbol{-k_{1}},-\boldsymbol{k_{1}-q}}g(k_{1}+q)]^2
\big)
\}
\nonumber \\
&-&
\beta \int\frac{d^{d}k}{(2\pi)^{d}} \left(\frac{3}{4}k^{4} \right)^2 \frac{e^{\beta k^{2}}g(k)}{(1+e^{\beta k^{2}})^{2}}\left[-i\omega g(k)+2\hbar\right]\:, \nonumber
\end{eqnarray}
where $K^{(conserving)}(\boldsymbol{k},\boldsymbol{k_1},\boldsymbol{q})$ describes the scattering amplitude of all collision processes that conserve the number of quasiparticles and holes. The Auger scattering processes are given by $K_{ph \to pp}$, $K_{hp\to pp}$, $K_{pp\to ph}$, and $K_{pp \to hp}$.
The explicit form of the scattering amplitude that conserve the particle/hole number is
\begin{eqnarray}
 K^{(conserving)}(\boldsymbol{k},\boldsymbol{k_1},\boldsymbol{q})
& =&
 \{V(\boldsymbol{q})^2 R_{++++}(\boldsymbol{k},\boldsymbol{k_1},\boldsymbol{k}-\boldsymbol{q},\boldsymbol{k}_1 + \boldsymbol{q})
 -
 V(\boldsymbol{q})V(\boldsymbol{k-q-k_{1}})
 Q_{++++}(\boldsymbol{k},\boldsymbol{k_1},\boldsymbol{k}-\boldsymbol{q},\boldsymbol{k}_1 + \boldsymbol{q}) \}
 \nonumber \\
 &-&
 \{
 V(\boldsymbol{k}+\boldsymbol{k}_1)^2
 R_{+--+}(\boldsymbol{k},-\boldsymbol{k_1}-\boldsymbol{q},-\boldsymbol{k}_1,\boldsymbol{k}-\boldsymbol{q})
 -
 V(\boldsymbol{k_{1}+k})V(\boldsymbol{q})
 Q_{+--+}(\boldsymbol{k},-\boldsymbol{k_1}-\boldsymbol{q},-\boldsymbol{k}_1,\boldsymbol{k}-\boldsymbol{q})
 \}
 \nonumber \\
& -&
 \{V(\boldsymbol{q})^2
 R_{++--}(\boldsymbol{k},-\boldsymbol{k}_1-\boldsymbol{q},\boldsymbol{k}-\boldsymbol{q},-\boldsymbol{k}_1)
 -
 V(\boldsymbol{q})V(\boldsymbol{k+k_{1}})
 Q_{++--}(\boldsymbol{k},-\boldsymbol{k}_1-\boldsymbol{q},\boldsymbol{k}-\boldsymbol{q},-\boldsymbol{k}_1)
 \}
 \:,
\end{eqnarray}
while the scattering amplitude describing the scattering process of a particle and a hole to two particles (holes) is given by:
\begin{eqnarray}
  K_{ph \to pp}(\boldsymbol{k},\boldsymbol{k}_1, \boldsymbol{q})
 &=&
 V(\boldsymbol{q})^2
 R_{++-+}(\boldsymbol{k},\boldsymbol{k_1},\boldsymbol{k}-\boldsymbol{q},\boldsymbol{k}_1 + \boldsymbol{q})
 -
 V(\boldsymbol{q}) V(\boldsymbol{k}-\boldsymbol{q}-\boldsymbol{k}_1)
 Q_{++-+}(\boldsymbol{k},\boldsymbol{k_1},\boldsymbol{k}-\boldsymbol{q},\boldsymbol{k}_1 + \boldsymbol{q})\\
 K_{hp\to pp}(\mathbf{k},\mathbf{k_1},\mathbf{q})
 &=&
 V(\boldsymbol{q})^2
 R_{-+++}(-\boldsymbol{k_1},-\boldsymbol{k},-\boldsymbol{k_1}-\boldsymbol{q},-\boldsymbol{k} + \boldsymbol{q})
 -
  V(\boldsymbol{q}) V(-\boldsymbol{k}-\boldsymbol{q}+\boldsymbol{k}_1)
  Q_{-+++}(-\boldsymbol{k_1},-\boldsymbol{k},-\boldsymbol{k_1}-\boldsymbol{q},-\boldsymbol{k} + \boldsymbol{q})\\
 K_{pp\to ph}(\mathbf{k},\mathbf{k_1},\mathbf{q})
 &=&
 V(\boldsymbol{q})^2
 R_{+++-}(-\boldsymbol{k}+\boldsymbol{q},\boldsymbol{k}_1,-\boldsymbol{k},\boldsymbol{k}_1+\boldsymbol{q})
 -
 V(\boldsymbol{q}) V(-\boldsymbol{k}-\boldsymbol{k_1})
 Q_{+++-}(-\boldsymbol{k}+\boldsymbol{q},\boldsymbol{k}_1,-\boldsymbol{k},\boldsymbol{k}_1+\boldsymbol{q}) \\
 K_{pp\to hp}(\mathbf{k},\mathbf{k_1},\mathbf{q})
 &=&
 V(\boldsymbol{q})^2
 R_{+-++}(-\boldsymbol{k}_1,\boldsymbol{k}-\boldsymbol{q},-\boldsymbol{k_1}-\boldsymbol{q},\boldsymbol{k})
 -
 V(\boldsymbol{q}) V(-\boldsymbol{k}-\boldsymbol{k_1})
 Q_{+-++}(-\boldsymbol{k}_1,\boldsymbol{k}-\boldsymbol{q},-\boldsymbol{k_1}-\boldsymbol{q},\boldsymbol{k})
 \:.
\end{eqnarray}
The problem of finding the stationary point of $\partial Q[g]/\partial g =0$ can be translated to the following matrix equation
\begin{equation}
 C^{coll}_{ab}\psi_b = \mathcal{L}_b
 \:,
\end{equation}
where $\mathcal{L}_b$ is given by
\begin{equation}
 \mathcal{L}_b
 =
 2 \hbar \beta \int\frac{d^{d}k}{(2\pi)^{d}} \left(\frac{3}{4}k^{4} \right)^2 \frac{e^{\beta k^{2}}}{(1+e^{\beta k^{2}})^{2}} L_b^{(3)}(k^2)
\end{equation}
and the collision matrix element has the form
\begin{eqnarray}
 C^{coll}_{ab}
 &=&
 -\beta(2\pi)^{1-3d}\int d^{d}kd^{d}k_{1}d^{d}q\,\{\frac{\delta(\epsilon_{\boldsymbol{k}}-\epsilon_{\boldsymbol{k}_{\boldsymbol{1}}+\boldsymbol{q}}-\epsilon_{\boldsymbol{k-q}}+\epsilon_{\boldsymbol{k_{1}}})K^{(conserving)}(\boldsymbol{k,k_{1,}q})}{(1+e^{\beta \epsilon_{\boldsymbol{k_{1}+q}}})(1+e^{\beta \epsilon_{\boldsymbol{k-q}}})(1+e^{-\beta \epsilon_{\boldsymbol{k}}})(1+e^{- \beta\epsilon_{k_{\boldsymbol{1}}}})}
 \nonumber \\
&& \qquad \quad \quad \times
\left[Y_{\boldsymbol{k,k_1}} L^{(3)}_a(k_{1}^2)+Y_{\boldsymbol{k,k}}L^{(3)}_a(k^2)-Y_{\boldsymbol{k,k_{1}+q}}L^{(3)}_a(|\boldsymbol{k_{1}}+\boldsymbol{q}|^2)-Y_{\boldsymbol{k,k-q}} L^{(3)}_a(|\boldsymbol{k}-\boldsymbol{q}|^3)\right]
\nonumber \\
&& \qquad \quad \quad \times
\left[Y_{\boldsymbol{k,k_1}} L^{(3)}_b(k_{1}^2)+Y_{\boldsymbol{k,k}}L^{(3)}_b(k^2)-Y_{\boldsymbol{k,k_{1}+q}}L^{(3)}_b(|\boldsymbol{k_{1}}+\boldsymbol{q}|^2)-Y_{\boldsymbol{k,k-q}} L^{(3)}_b(|\boldsymbol{k}-\boldsymbol{q}|^3)\right]
\nonumber \\
&&
\qquad \quad +
\frac{\delta(\epsilon_{\boldsymbol{k}}-\epsilon_{\boldsymbol{k}_{1}}-\epsilon_{\boldsymbol{k-q}}-\epsilon_{\boldsymbol{k_{1}+q}}) K_{ph\to pp}(\boldsymbol{k},\boldsymbol{k_1},\boldsymbol{q}) }{(1+e^{\beta \epsilon_{\boldsymbol{k_{1}+q}}})(1+e^{\beta \epsilon_{\boldsymbol{k-q}}})(1+e^{-\beta \epsilon_{\boldsymbol{k}}})(1+e^{\beta \epsilon_{\boldsymbol{k}_{1}}})}\\
&& \qquad \quad \quad
\times
\left[ -Y_{\boldsymbol{k},\boldsymbol{k}_{1}} L^{(3)}_a(k_{1}^2)+Y_{\boldsymbol{k,k}}L^{(3)}_a(k^2)-Y_{\boldsymbol{k},\boldsymbol{k_{1}+q}} L^{(3)}_a(|\boldsymbol{k_{1}} + \boldsymbol{q}|^2)-Y_{\boldsymbol{k},\boldsymbol{k-q}} L^{(3)}_a(\ \boldsymbol{k} - \boldsymbol{q}) \right]
\nonumber \\
&& \qquad \quad \quad
\times
\left[ -Y_{\boldsymbol{k},\boldsymbol{k}_{1}} L^{(3)}_b(k_{1}^2)+Y_{\boldsymbol{k,k}}L^{(3)}_b(k^2)-Y_{\boldsymbol{k},\boldsymbol{k_{1}+q}} L^{(3)}_b(|\boldsymbol{k_{1}} + \boldsymbol{q}|^2)-Y_{\boldsymbol{k},\boldsymbol{k-q}} L^{(3)}_b(\ \boldsymbol{k} - \boldsymbol{q}) \right]
\nonumber \\
&& \qquad \quad
+ K_{hp\to pp}\cdots +K_{pp\to hp} \cdots + K_{pp \to ph} \cdots
\}
\:. \nonumber
\end{eqnarray}
After having determined the elements of the collision matrix and the vector $\mathcal{L}_b$, the coefficient $\psi_b$ which enter the distribution function are given by an inversion of the matrix equation with
\begin{equation}
 \psi_a= \left(C^{coll}_{ab} \right)^{-1} \mathcal{L}_b
 \:.
\end{equation}
The shear viscosity of Luttinger semimetals can be evaluated by the expression
\begin{equation}
 \eta= \beta \sum_{n,\lambda= \pm 1} \int \frac{d^d k}{(2 \pi)^d} v^{\alpha}_{\boldsymbol{k}} k^{\beta} I_{\alpha \beta} f^{(0)}_{\lambda}[1-f^{(0)}_{\lambda}] \psi_n L^{(3)}_n(k^2)
 \:,
\end{equation}
(where we do not sum over the indices $\alpha$ and $\beta$, and $\alpha \neq \beta$.) In the determination of the shear viscosity, we took the Laguerre-polynomials $L^{(3)}_n$ with $n \in \{0,1,2,3,4,5\}$ into account and checked that the sum over $n$ converges.
%
%
%
%
%
\subsection{Electrical conductivity}
In this section, the electrical conductivity of Luttinger semi-metals is determined. The procedure is analogous to the calculation of the viscosity. However, this time we assume that the distribution function is spatially homogeneous, i.e. $f_{\lambda}$ does not depend on the spatial coordinate, and that the distribution function is independent of the time $t$. The ansatz for the out-of-equilibrium distribution function is given by
$h_{\sigma}(\boldsymbol{k},\lambda)=\lambda v_{\boldsymbol{k}}^{\alpha} E^{\alpha}$, where $E^{\alpha}$ is the $\alpha$- component of the electrical field. For the Boltzmann equation, we find the expression:
\begin{equation}
 -e v^{\alpha}_k\lambda \beta f^{(0)}_{\lambda}(1-f^{(0)}_{\lambda}) E^{\alpha}
 =
 I^{lin,\sigma}_{\lambda}(\boldsymbol{k})
\end{equation}
with the collision integral:
%
\begin{align}
 I^{lin,\sigma}_{\lambda}(\boldsymbol{k})
 &=\beta(2\pi)^{d+1-3d}\int d^{d}k_{1}d^{d}q\,\delta(\mu\epsilon_{\boldsymbol{k}}+\lambda\epsilon_{\boldsymbol{k}_{\boldsymbol{1}}}-\rho\epsilon_{\boldsymbol{k-q}}-\nu\epsilon_{\boldsymbol{k_{1}+q}})\:E^{\alpha}
 \nonumber \\
 \times&
 \{V(\boldsymbol{q})^2 R_{\mu \rho \lambda \nu}(\boldsymbol{k},\boldsymbol{k_1},\boldsymbol{k-q},\boldsymbol{k_1}+\boldsymbol{q})
 -
 V(\boldsymbol{q})V(\boldsymbol{k-q-k_{1}}) Q_{\mu \rho \lambda \nu}(\boldsymbol{k},\boldsymbol{k_1},\boldsymbol{k-q},\boldsymbol{k_1}+\boldsymbol{q}) \}
 \\
 \times&\frac{\lambda v_{k_{1}}^{\alpha}g(k_{1})+\mu v_{k}^{\alpha}g(k)-\nu v_{k_{1}+q}^{\alpha}g(k_{1}+q)-\rho v_{k-q}^{\alpha}g(k-q)}{(1+e^{\beta \nu\epsilon_{\boldsymbol{k_{1}+q}}})(1+e^{\beta \rho\epsilon_{\boldsymbol{k-q}}})(1+e^{-\beta \mu\epsilon_{\boldsymbol{k}}})(1+e^{-\beta \lambda\epsilon_{\boldsymbol{k}_{\boldsymbol{1}}}})}
 \:.
 \nonumber
\end{align}
%
Next we introduce the function
\begin{equation}
 R_{\boldsymbol{k},\boldsymbol{k}_1}
 =
 \boldsymbol{v_{k}}\cdot \boldsymbol{v_{k_1}}
 =
 4 \boldsymbol{k}\cdot \boldsymbol{k_1}
 \:
\end{equation}
into the Boltzmann equation. The collision operator $I_{coll}[\tilde{h}_{\sigma}]$ with $\tilde{h}_{\sigma}=\lambda R_{\boldsymbol{k},\boldsymbol{k}}g(k)$ is again self adjoint with respect to the scalar product and the following functional is found
%
\begin{eqnarray}
&& Q_{\sigma}[g]= \beta \frac{(2\pi)^{d+1-3d}}{2}\int d^{d}k d^{d}k_{1}d^{d}q
\{
\frac{\delta(\epsilon_{\boldsymbol{k}}+\epsilon_{\boldsymbol{k}_{1}}-\epsilon_{\boldsymbol{k-q}}-\epsilon_{\boldsymbol{k_{1}+q}}) T^{(1)}(\boldsymbol{k},\boldsymbol{k_1},\boldsymbol{q})}{(1+e^{\beta \epsilon_{\boldsymbol{k_{1}+q}}})(1+e^{\beta \epsilon_{\boldsymbol{k-q}}})(1+e^{-\beta \epsilon_{\boldsymbol{k}}})(1+e^{-\beta \epsilon_{\boldsymbol{k}_{\boldsymbol{1}}}})}
\nonumber \\
&& \qquad \qquad \quad
\times \left[R_{k,k_{1}}g(k_{1})+R_{k,k}g(k)-R_{k,k_{1}+q}g(k_{1}+q)-R_{k,k-q}g(k-q) \right]^2
\nonumber \\
&& \qquad \qquad
+
\frac{\delta(\epsilon_{\boldsymbol{k}}+\epsilon_{\boldsymbol{k}_{\boldsymbol{1}}}-\epsilon_{\boldsymbol{k-q}}-\epsilon_{\boldsymbol{k_{1}+q}}) T^{(2)}(\boldsymbol{k},\boldsymbol{k_1},\boldsymbol{q})}{(1+e^{\beta \epsilon_{\boldsymbol{k_{1}+q}}})(1+e^{\beta \epsilon_{\boldsymbol{k-q}}})(1+e^{-\beta \epsilon_{\boldsymbol{k}}})(1+e^{-\beta \epsilon_{\boldsymbol{k}_{\boldsymbol{1}}}})}
\nonumber \\
&& \qquad \qquad \quad
\times
\left[R_{k,k}g(k)+R_{k,k_{1}+q}g(k_{1}+q)-R_{k,k-q}g(k-q)-R_{k,k_{1}}g(k_{1}) \right]^2
\nonumber \\
&& \qquad \qquad
+
\frac{\delta(\epsilon_{\boldsymbol{k}}-\epsilon_{\boldsymbol{k}_{1}}-\epsilon_{\boldsymbol{k-q}}-\epsilon_{\boldsymbol{k_{1}+q}}) }{(1+e^{\beta \epsilon_{\boldsymbol{k_{1}+q}}})(1+e^{\beta \epsilon_{\boldsymbol{k-q}}})(1+e^{- \beta \epsilon_{\boldsymbol{k}}})(1+e^{\beta \epsilon_{\boldsymbol{k}_{1}}})}
\nonumber \\
&& \qquad \qquad \quad
\times
\big(
K_{ph \to pp}(\boldsymbol{k},\boldsymbol{k_1},\boldsymbol{q})
\left[ -R_{\boldsymbol{k},\boldsymbol{k}_{1}}g(k_{1})+R_{\boldsymbol{k,k}}g(k)-R_{\boldsymbol{k},\boldsymbol{k_{1}+q}}g(k_{1}+q)-R_{\boldsymbol{k},\boldsymbol{k-q}}g(k-q) \right]^2
\nonumber \\
&& \qquad \qquad \qquad
+K_{hp \to pp}(\boldsymbol{k},\boldsymbol{k_1},\boldsymbol{q})
[+ R_{k,k_{1}}g(k)- R_{k_{1},k_{1}}g(k_{1})-R_{k_{1},k-q}g(k-q)- R_{k_{1},k_{1}+q}g(k_{1}+q)]^2\\
&& \qquad \qquad \qquad
-
K_{pp\to ph}(\boldsymbol{k},\boldsymbol{k_1},\boldsymbol{q})
[- R_{k-q,k_{1}}g(k_{1})+ R_{k-q,k-q}g(k-q)- R_{k-q,k_{1}+q}g(k_{1}+q)- R_{k-q,k}g(k)]^2
\nonumber \\
&& \qquad \qquad \qquad
-
K_{pp \to hp}(\boldsymbol{k},\boldsymbol{k_1},\boldsymbol{q})
[- R_{k_{1},k-q}g(k-q)+ R_{k_{1},k_{1}}g(k_{1})+ R_{k_{1},k}g(k)+ R_{k_{1},k_{1}+q}g(k_{1}+q)]^2
\big)
\}
\nonumber \\
&& \qquad
+
\beta \int\frac{d^{d}k}{(2\pi)^{d}} (R_{\boldsymbol{k},\boldsymbol{k}})^2\,\frac{e^{\beta k^{2}}}{(1+e^{\beta k^{2}})^{2}} g(k)
\:.
\end{eqnarray}
The intraband scattering processes are determined by the scattering amplitude $T^{(1)}$ with
\begin{equation}
 T^{(1)}(\boldsymbol{k},\boldsymbol{k_1},\boldsymbol{q})
 =
 V(\boldsymbol{q})^2 R_{++++}(\boldsymbol{k},\boldsymbol{k_1},\boldsymbol{k}-\boldsymbol{q},\boldsymbol{k}_1 + \boldsymbol{q})
 -
 V(\boldsymbol{q})V(\boldsymbol{k-q-k_{1}})
 Q_{++++}(\boldsymbol{k},\boldsymbol{k_1},\boldsymbol{k}-\boldsymbol{q},\boldsymbol{k}_1 + \boldsymbol{q})
 \:,
\end{equation}
while the interband scattering processes that conserve the particle and hole number have the following scattering amplitude:
\begin{align}
 -T^{(2)}(\boldsymbol{k},\boldsymbol{k_1},\boldsymbol{q})
 &=
 V(\boldsymbol{k}+\boldsymbol{k}_1)^2
 R_{+--+}(\boldsymbol{k},-\boldsymbol{k_1}-\boldsymbol{q},-\boldsymbol{k}_1,\boldsymbol{k}-\boldsymbol{q})
 -
 V(\boldsymbol{k_{1}+k})V(\boldsymbol{q})
 Q_{+--+}(\boldsymbol{k},-\boldsymbol{k_1}-\boldsymbol{q},-\boldsymbol{k}_1,\boldsymbol{k}-\boldsymbol{q})
 \nonumber \\
& +
 \{V(\boldsymbol{q})^2
 R_{++--}(\boldsymbol{k},-\boldsymbol{k}_1-\boldsymbol{q},\boldsymbol{k}-\boldsymbol{q},-\boldsymbol{k}_1)
 -
 V(\boldsymbol{q})V(\boldsymbol{k+k_{1}})
 Q_{++--}(\boldsymbol{k},-\boldsymbol{k}_1-\boldsymbol{q},\boldsymbol{k}-\boldsymbol{q},-\boldsymbol{k}_1)
 \}
 \:.
\end{align}
Similar to the calculation of the viscosity, the coefficients $\psi_n$ of the distribution function are determined by inverting the matrix equation arising from the above functional $Q_{\sigma}[g]$. The electrical conductivity can be evaluated by using
\begin{equation}
 \sigma_{\alpha \beta}
 =
 \beta \sum_{n,\lambda=\pm 1}
 v^{\alpha}_{\boldsymbol{k}} v^{\beta}_{\boldsymbol{k}} f^{(0)}_{\lambda}[1-f^{(0)}_{\lambda}] \psi_n L^{(3)}_n(k^2)
 \:.
\end{equation}
For the result given in the paper, the Laguerre polynomials with $n$ from 0 to 4 were taken into account.
\section{Condition on the transferred momentum set by energy conservation}
In this section, we elaborate further on how to evaluate the integrals. Especially, we focus on the conditions for the momenta enforced by the $\delta$-distribution of the energy conservation. Here, we have to distinguish between the two different classes of scattering processes, namely the scattering processes that conserve the particle/hole number and those that violate it.

 Let us first start with the scattering processes that conserve the particle and hole number, and focus on the intraband scattering case, since by shifting the variables  the other two interband processes preserving particle/hole number can be cast into a form with the same $\delta$ function.

 For the intraband scattering processes, the energy conservation is implemented by $\delta(\epsilon_{\boldsymbol{k}}+\epsilon_{\boldsymbol{k}_{1}}-\epsilon_{\boldsymbol{k-q}}-\epsilon_{\boldsymbol{k_{1}+q}})\,$, which corresponds to
 $|\boldsymbol{k}|^{2}+|\boldsymbol{k_{1}}|^{2}=	|\boldsymbol{k}-\boldsymbol{q}|^{2}+|\boldsymbol{k_{1}+q}|^{2}\:.$  Upon using the parametrization $\boldsymbol{q=}\frac{1}{2}(\boldsymbol{k-}\boldsymbol{k}_{1})+\boldsymbol{Q}$, the $\delta$-distribution of the energy conservation
 can be cast into the following form
 %
 \begin{equation}
  \delta(\epsilon_{\boldsymbol{k}}+\epsilon_{\boldsymbol{k}_{1}}-\epsilon_{\boldsymbol{k-q}}-\epsilon_{\boldsymbol{k_{1}+q}})=	 \delta\left(\frac{1}{4}|\boldsymbol{k}-\boldsymbol{k_{1}}|^{2}-Q^{2}\right)
=	 \frac{1}{|\boldsymbol{k}-\boldsymbol{k_{1}}|}\left[\delta\left(\frac{1}{2}|\boldsymbol{k}-\boldsymbol{k_{1}}|-Q\right)+\delta\left(\frac{1}{2}|\boldsymbol{k}-\boldsymbol{k_{1}}|+Q\right)\right]
\:.
 \end{equation}
%
%
The consequence of this relation is that the momentum $\boldsymbol{q}$ which is transferred during the scattering processes between the colliding particles/holes is restricted to
\begin{equation}
 \boldsymbol{q=}\frac{1}{2}(\boldsymbol{k-}\boldsymbol{k}_{1})\pm\frac{1}{2}|\boldsymbol{k}-\boldsymbol{k_{1}}|\hat{e}_{r}
 \:,
\end{equation}
where $\hat{e}_{r}$ is the radial unit vector in $d$-dimensions.

Next we consider the scattering processes which do not conserve the particle and hole number, such as ${e^-_{\boldsymbol{k}} h^+_{\boldsymbol{k_1}} \to e^-_{\boldsymbol{k}-\boldsymbol{q}} e^-_{\boldsymbol{k_1}+\boldsymbol{q}}}$. The energy conservation implies that
\begin{equation}
 \boldsymbol{k}^{2}-\boldsymbol{k_1}^{2}
 =
 |\boldsymbol{k-q}|^{2}+|\boldsymbol{k_{1}}+\boldsymbol{q}|^{2}
 \:.
\end{equation}
Again upon introducing the following parametrization for $\boldsymbol{q}=\frac{1}{2}(\boldsymbol{k}-\boldsymbol{k}_{1})+\boldsymbol{Q}$, it holds:
%
 \begin{eqnarray}
  \delta(\epsilon_{\boldsymbol{k}}-\epsilon_{\boldsymbol{k}_{1}}-\epsilon_{\boldsymbol{k-q}}-\epsilon_{\boldsymbol{k_{1}+q}})
  &=&
  \delta\left(\frac{1}{4}|\boldsymbol{k}-\boldsymbol{k_{1}}|^{2}-k_{1}^{2}-Q^{2}\right) \\
&=&	 \frac{1}{2\sqrt{\frac{|\boldsymbol{k}-\boldsymbol{k}_{1}|^{2}}{4}-k_{1}^{2}}}\left[\delta\left(\sqrt{\frac{|\boldsymbol{k}-\boldsymbol{k}_{1}|{}^{2}}{4}-k_{1}^{2}}-Q\right)+\delta\left(\sqrt{\frac{|\boldsymbol{k}-\boldsymbol{k}_{1}|{}^{2}}{4}-k_{1}^{2}}+q\right)\right]\:,
 \end{eqnarray}
where the absolute value $k$ is restricted to
\begin{equation}
 k
 \ge
 k_{1}\left(\cos\theta_{kk_{1}}+\sqrt{3+\cos^{2}\theta_{kk_{1}}}\right)
 \:.
 \label{eq:lowerbound_k}
\end{equation}
$\theta_{kk_1}$ is the angle between the two momentum $\boldsymbol{k}$ and $\boldsymbol{k}_1$.
The $\boldsymbol{q}$-momentum for non-conserving particle number is then given by $\boldsymbol{q}=\frac{1}{2}(\boldsymbol{k-}\boldsymbol{k}_{1})\pm\sqrt{\frac{|\boldsymbol{k}-\boldsymbol{k}_{1}|{}^{2}}{4}-\boldsymbol{k}_{1}^{2}}\hat{e_{r}\:,}$ where $\hat{e_{r}}$ is the radial unit vector in $d$-dimensions.
Equation~\eqref{eq:lowerbound_k} sets the lower limit in the integration of the absolute value of $\boldsymbol{k}$.
\end{widetext}


\begin{thebibliography}{30}%
\makeatletter
\providecommand \@ifxundefined [1]{%
 \@ifx{#1\undefined}
}%
\providecommand \@ifnum [1]{%
 \ifnum #1\expandafter \@firstoftwo
 \else \expandafter \@secondoftwo
 \fi
}%
\providecommand \@ifx [1]{%
 \ifx #1\expandafter \@firstoftwo
 \else \expandafter \@secondoftwo
 \fi
}%
\providecommand \natexlab [1]{#1}%
\providecommand \enquote  [1]{``#1''}%
\providecommand \bibnamefont  [1]{#1}%
\providecommand \bibfnamefont [1]{#1}%
\providecommand \citenamefont [1]{#1}%
\providecommand \href@noop [0]{\@secondoftwo}%
\providecommand \href [0]{\begingroup \@sanitize@url \@href}%
\providecommand \@href[1]{\@@startlink{#1}\@@href}%
\providecommand \@@href[1]{\endgroup#1\@@endlink}%
\providecommand \@sanitize@url [0]{\catcode `\\12\catcode `\$12\catcode
  `\&12\catcode `\#12\catcode `\^12\catcode `\_12\catcode `\%12\relax}%
\providecommand \@@startlink[1]{}%
\providecommand \@@endlink[0]{}%
\providecommand \url  [0]{\begingroup\@sanitize@url \@url }%
\providecommand \@url [1]{\endgroup\@href {#1}{\urlprefix }}%
\providecommand \urlprefix  [0]{URL }%
\providecommand \Eprint [0]{\href }%
\providecommand \doibase [0]{http://dx.doi.org/}%
\providecommand \selectlanguage [0]{\@gobble}%
\providecommand \bibinfo  [0]{\@secondoftwo}%
\providecommand \bibfield  [0]{\@secondoftwo}%
\providecommand \translation [1]{[#1]}%
\providecommand \BibitemOpen [0]{}%
\providecommand \bibitemStop [0]{}%
\providecommand \bibitemNoStop [0]{.\EOS\space}%
\providecommand \EOS [0]{\spacefactor3000\relax}%
\providecommand \BibitemShut  [1]{\csname bibitem#1\endcsname}%
\let\auto@bib@innerbib\@empty
\bibitem [{\citenamefont {Damle}\ and\ \citenamefont
  {Sachdev}(1997)}]{Damle1997}%
  \BibitemOpen
  \bibfield  {author} {\bibinfo {author} {\bibfnamefont {K.}~\bibnamefont
  {Damle}}\ and\ \bibinfo {author} {\bibfnamefont {S.}~\bibnamefont
  {Sachdev}},\ }\href {\doibase 10.1103/physrevb.56.8714} {\bibfield  {journal}
  {\bibinfo  {journal} {Phys. Rev. B}\ }\textbf {\bibinfo {volume} {56}},\
  \bibinfo {pages} {8714} (\bibinfo {year} {1997})}\BibitemShut {NoStop}%
\bibitem [{\citenamefont {Levitov}\ and\ \citenamefont
  {Falkovich}(2016)}]{Levitov2016}%
  \BibitemOpen
  \bibfield  {author} {\bibinfo {author} {\bibfnamefont {L.}~\bibnamefont
  {Levitov}}\ and\ \bibinfo {author} {\bibfnamefont {G.}~\bibnamefont
  {Falkovich}},\ }\href {\doibase 10.1038/nphys3667} {\bibfield  {journal}
  {\bibinfo  {journal} {Nat. Phys.}\ }\textbf {\bibinfo {volume} {12}},\
  \bibinfo {pages} {672} (\bibinfo {year} {2016})}\BibitemShut {NoStop}%
\bibitem [{\citenamefont {{\hspace{0.167em}}K. Kovtun}\ \emph
  {et~al.}(2005)\citenamefont {{\hspace{0.167em}}K. Kovtun}, \citenamefont
  {Son},\ and\ \citenamefont {Starinets}}]{Kovtun2005}%
  \BibitemOpen
  \bibfield  {author} {\bibinfo {author} {\bibfnamefont {P.~K.}~\bibnamefont
  {Kovtun}}, \bibinfo {author} {\bibfnamefont {D.~T.}\
  \bibnamefont {Son}}, \ and\ \bibinfo {author} {\bibfnamefont {A.~O.}\
  \bibnamefont {Starinets}},\ }\href {\doibase 10.1103/physrevlett.94.111601}
  {\bibfield  {journal} {\bibinfo  {journal} {Phys. Rev. Lett.}\ }\textbf
  {\bibinfo {volume} {94}},\ \bibinfo {pages} {111601} (\bibinfo {year}
  {2005})}\BibitemShut {NoStop}%
\bibitem [{\citenamefont {Kovtun}\ and\ \citenamefont
  {Nickel}(2009)}]{Kovtun2009}%
  \BibitemOpen
  \bibfield  {author} {\bibinfo {author} {\bibfnamefont {P.}~\bibnamefont
  {Kovtun}}\ and\ \bibinfo {author} {\bibfnamefont {D.}~\bibnamefont
  {Nickel}},\ }\href {\doibase 10.1103/physrevlett.102.011602} {\bibfield
  {journal} {\bibinfo  {journal} {Phys. Rev. Lett.}\ }\textbf {\bibinfo
  {volume} {102}},\ \bibinfo {pages} {011602} (\bibinfo {year}
  {2009})}\BibitemShut {NoStop}%
\bibitem [{\citenamefont {Narozhny}\ \emph {et~al.}(2017)\citenamefont
  {Narozhny}, \citenamefont {Gornyi}, \citenamefont {Mirlin},\ and\
  \citenamefont {Schmalian}}]{Narozhny2017}%
  \BibitemOpen
  \bibfield  {author} {\bibinfo {author} {\bibfnamefont {B.~N.}\ \bibnamefont
  {Narozhny}}, \bibinfo {author} {\bibfnamefont {I.~V.}\ \bibnamefont
  {Gornyi}}, \bibinfo {author} {\bibfnamefont {A.~D.}\ \bibnamefont {Mirlin}},
  \ and\ \bibinfo {author} {\bibfnamefont {J.}~\bibnamefont {Schmalian}},\
  }\href {\doibase 10.1002/andp.201700043} {\bibfield  {journal} {\bibinfo
  {journal} {Annalen der Physik}\ }\textbf {\bibinfo {volume} {529}},\ \bibinfo
  {pages} {1700043} (\bibinfo {year} {2017})}\BibitemShut {NoStop}%
\bibitem [{\citenamefont {Lucas}\ and\ \citenamefont {Fong}(2018)}]{Lucas2018}%
  \BibitemOpen
  \bibfield  {author} {\bibinfo {author} {\bibfnamefont {A.}~\bibnamefont
  {Lucas}}\ and\ \bibinfo {author} {\bibfnamefont {K.~C.}\ \bibnamefont
  {Fong}},\ }\href {\doibase 10.1088/1361-648x/aaa274} {\bibfield  {journal}
  {\bibinfo  {journal} {J. Phys.: Condens. Matter}\ }\textbf
  {\bibinfo {volume} {30}},\ \bibinfo {pages} {053001} (\bibinfo {year}
  {2018})}\BibitemShut {NoStop}%
\bibitem [{\citenamefont {Crossno}\ \emph {et~al.}(2016)\citenamefont
  {Crossno}, \citenamefont {Shi}, \citenamefont {Wang}, \citenamefont {Liu},
  \citenamefont {Harzheim}, \citenamefont {Lucas}, \citenamefont {Sachdev},
  \citenamefont {Kim}, \citenamefont {Taniguchi}, \citenamefont {Watanabe},
  \citenamefont {Ohki},\ and\ \citenamefont {Fong}}]{Crossno2016}%
  \BibitemOpen
  \bibfield  {author} {\bibinfo {author} {\bibfnamefont {J.}~\bibnamefont
  {Crossno}}, \bibinfo {author} {\bibfnamefont {J.~K.}\ \bibnamefont {Shi}},
  \bibinfo {author} {\bibfnamefont {K.}~\bibnamefont {Wang}}, \bibinfo {author}
  {\bibfnamefont {X.}~\bibnamefont {Liu}}, \bibinfo {author} {\bibfnamefont
  {A.}~\bibnamefont {Harzheim}}, \bibinfo {author} {\bibfnamefont
  {A.}~\bibnamefont {Lucas}}, \bibinfo {author} {\bibfnamefont
  {S.}~\bibnamefont {Sachdev}}, \bibinfo {author} {\bibfnamefont
  {P.}~\bibnamefont {Kim}}, \bibinfo {author} {\bibfnamefont {T.}~\bibnamefont
  {Taniguchi}}, \bibinfo {author} {\bibfnamefont {K.}~\bibnamefont {Watanabe}},
  \bibinfo {author} {\bibfnamefont {T.~A.}\ \bibnamefont {Ohki}}, \ and\
  \bibinfo {author} {\bibfnamefont {K.~C.}\ \bibnamefont {Fong}},\ }\href
  {\doibase 10.1126/science.aad0343} {\bibfield  {journal} {\bibinfo  {journal}
  {Science}\ }\textbf {\bibinfo {volume} {351}},\ \bibinfo {pages} {1058}
  (\bibinfo {year} {2016})}\BibitemShut {NoStop}%
\bibitem [{\citenamefont {M{\"u}ller}\ \emph {et~al.}(2009)\citenamefont
  {M{\"u}ller}, \citenamefont {Schmalian},\ and\ \citenamefont
  {Fritz}}]{Mueller2009}%
  \BibitemOpen
  \bibfield  {author} {\bibinfo {author} {\bibfnamefont {M.}~\bibnamefont
  {M{\"u}ller}}, \bibinfo {author} {\bibfnamefont {J.}~\bibnamefont
  {Schmalian}}, \ and\ \bibinfo {author} {\bibfnamefont {L.}~\bibnamefont
  {Fritz}},\ }\href {\doibase 10.1103/physrevlett.103.025301} {\bibfield
  {journal} {\bibinfo  {journal} {Phys. Rev. Lett.}\ }\textbf {\bibinfo
  {volume} {103}},\ \bibinfo {pages} {025301} (\bibinfo {year}
  {2009})}\BibitemShut {NoStop}%
\bibitem [{\citenamefont {Herbut}(2009)}]{Herbut2009}%
  \BibitemOpen
  \bibfield  {author} {\bibinfo {author} {\bibfnamefont {I.}~\bibfnamefont {F.}~\bibnamefont
  {Herbut}},\ }\href {\doibase 10.1103/physics.2.57} {\bibfield  {journal}
  {\bibinfo  {journal} {Physics}\ }\textbf {\bibinfo {volume} {2}},\ \bibinfo
  {pages} {57} (\bibinfo {year} {2009})}\BibitemShut {NoStop}%
\bibitem [{\citenamefont {Fritz}\ \emph {et~al.}(2008)\citenamefont {Fritz},
  \citenamefont {Schmalian}, \citenamefont {M{\"u}ller},\ and\ \citenamefont
  {Sachdev}}]{Fritz2008}%
  \BibitemOpen
  \bibfield  {author} {\bibinfo {author} {\bibfnamefont {L.}~\bibnamefont
  {Fritz}}, \bibinfo {author} {\bibfnamefont {J.}~\bibnamefont {Schmalian}},
  \bibinfo {author} {\bibfnamefont {M.}~\bibnamefont {M{\"u}ller}}, \ and\
  \bibinfo {author} {\bibfnamefont {S.}~\bibnamefont {Sachdev}},\ }\href
  {\doibase 10.1103/physrevb.78.085416} {\bibfield  {journal} {\bibinfo
  {journal} {Phys. Rev. B}\ }\textbf {\bibinfo {volume} {78}},\ \bibinfo
  {pages} {085416} (\bibinfo {year} {2008})}\BibitemShut {NoStop}%
\bibitem [{\citenamefont {Bandurin}\ \emph {et~al.}(2016)\citenamefont
  {Bandurin}, \citenamefont {Torre}, \citenamefont {Kumar}, \citenamefont
  {Shalom}, \citenamefont {Tomadin}, \citenamefont {Principi}, \citenamefont
  {Auton}, \citenamefont {Khestanova}, \citenamefont {Novoselov}, \citenamefont
  {Grigorieva}, \citenamefont {Ponomarenko}, \citenamefont {Geim},\ and\
  \citenamefont {Polini}}]{Bandurin2016}%
  \BibitemOpen
  \bibfield  {author} {\bibinfo {author} {\bibfnamefont {D.~A.}\ \bibnamefont
  {Bandurin}}, \bibinfo {author} {\bibfnamefont {I.}~\bibnamefont {Torre}},
  \bibinfo {author} {\bibfnamefont {R.~K.}\ \bibnamefont {Kumar}}, \bibinfo
  {author} {\bibfnamefont {M.~B.}\ \bibnamefont {Shalom}}, \bibinfo {author}
  {\bibfnamefont {A.}~\bibnamefont {Tomadin}}, \bibinfo {author} {\bibfnamefont
  {A.}~\bibnamefont {Principi}}, \bibinfo {author} {\bibfnamefont {G.~H.}\
  \bibnamefont {Auton}}, \bibinfo {author} {\bibfnamefont {E.}~\bibnamefont
  {Khestanova}}, \bibinfo {author} {\bibfnamefont {K.~S.}\ \bibnamefont
  {Novoselov}}, \bibinfo {author} {\bibfnamefont {I.~V.}\ \bibnamefont
  {Grigorieva}}, \bibinfo {author} {\bibfnamefont {L.~A.}\ \bibnamefont
  {Ponomarenko}}, \bibinfo {author} {\bibfnamefont {A.~K.}\ \bibnamefont
  {Geim}}, \ and\ \bibinfo {author} {\bibfnamefont {M.}~\bibnamefont
  {Polini}},\ }\href {\doibase 10.1126/science.aad0201} {\bibfield  {journal}
  {\bibinfo  {journal} {Science}\ }\textbf {\bibinfo {volume} {351}},\ \bibinfo
  {pages} {1055} (\bibinfo {year} {2016})}\BibitemShut {NoStop}%
\bibitem [{\citenamefont {Titov}\ \emph {et~al.}(2013)\citenamefont {Titov},
  \citenamefont {Gorbachev}, \citenamefont {Narozhny}, \citenamefont
  {Tudorovskiy}, \citenamefont {Sch{\"u}tt}, \citenamefont {Ostrovsky},
  \citenamefont {Gornyi}, \citenamefont {Mirlin}, \citenamefont {Katsnelson},
  \citenamefont {Novoselov}, \citenamefont {Geim},\ and\ \citenamefont
  {Ponomarenko}}]{Titov2013}%
  \BibitemOpen
  \bibfield  {author} {\bibinfo {author} {\bibfnamefont {M.}~\bibnamefont
  {Titov}}, \bibinfo {author} {\bibfnamefont {R.~V.}\ \bibnamefont
  {Gorbachev}}, \bibinfo {author} {\bibfnamefont {B.~N.}\ \bibnamefont
  {Narozhny}}, \bibinfo {author} {\bibfnamefont {T.}~\bibnamefont
  {Tudorovskiy}}, \bibinfo {author} {\bibfnamefont {M.}~\bibnamefont
  {Sch{\"u}tt}}, \bibinfo {author} {\bibfnamefont {P.~M.}\ \bibnamefont
  {Ostrovsky}}, \bibinfo {author} {\bibfnamefont {I.~V.}\ \bibnamefont
  {Gornyi}}, \bibinfo {author} {\bibfnamefont {A.~D.}\ \bibnamefont {Mirlin}},
  \bibinfo {author} {\bibfnamefont {M.~I.}\ \bibnamefont {Katsnelson}},
  \bibinfo {author} {\bibfnamefont {K.~S.}\ \bibnamefont {Novoselov}}, \bibinfo
  {author} {\bibfnamefont {A.~K.}\ \bibnamefont {Geim}}, \ and\ \bibinfo
  {author} {\bibfnamefont {L.~A.}\ \bibnamefont {Ponomarenko}},\ }\href
  {\doibase 10.1103/physrevlett.111.166601} {\bibfield  {journal} {\bibinfo
  {journal} {Phys. Rev. Lett.}\ }\textbf {\bibinfo {volume} {111}},\ \bibinfo
  {pages} {166601} (\bibinfo {year} {2013})}\BibitemShut {NoStop}%
\bibitem [{\citenamefont {Link}\ \emph
  {et~al.}(2018{\natexlab{a}})\citenamefont {Link}, \citenamefont {Narozhny},
  \citenamefont {Kiselev},\ and\ \citenamefont {Schmalian}}]{Link2018}%
  \BibitemOpen
  \bibfield  {author} {\bibinfo {author} {\bibfnamefont {J.~M.}\ \bibnamefont
  {Link}}, \bibinfo {author} {\bibfnamefont {B.~N.}\ \bibnamefont {Narozhny}},
  \bibinfo {author} {\bibfnamefont {E.~I.}\ \bibnamefont {Kiselev}}, \ and\
  \bibinfo {author} {\bibfnamefont {J.}~\bibnamefont {Schmalian}},\ }\href
  {\doibase 10.1103/physrevlett.120.196801} {\bibfield  {journal} {\bibinfo
  {journal} {Phys. Rev. Lett.}\ }\textbf {\bibinfo {volume} {120}},\ \bibinfo
  {pages} {196801} (\bibinfo {year} {2018}{\natexlab{a}})}\BibitemShut
  {NoStop}%
\bibitem [{\citenamefont {Luttinger}(1956)}]{Luttinger}%
  \BibitemOpen
  \bibfield  {author} {\bibinfo {author} {\bibfnamefont {J.~M.}\ \bibnamefont
  {Luttinger}},\ }\href {\doibase 10.1103/physrev.102.1030} {\bibfield
  {journal} {\bibinfo  {journal} {Phys. Rev.}\ }\textbf {\bibinfo {volume}
  {102}},\ \bibinfo {pages} {1030} (\bibinfo {year} {1956})}\BibitemShut
  {NoStop}%
\bibitem [{\citenamefont {Abrikosov}(1974)}]{Abrikosov1974}%
  \BibitemOpen
  \bibfield  {author} {\bibinfo {author} {\bibfnamefont {A.}~\bibfnamefont {A.}~\bibnamefont
  {Abrikosov}},\ }\href@noop {} {\bibfield  {journal} {\bibinfo  {journal}
  {Sov. Phys. JETP}\ }\textbf {\bibinfo {volume} {39}},\ \bibinfo {pages} {709}
  (\bibinfo {year} {1974})}\BibitemShut {NoStop}%
\bibitem [{\citenamefont {Wilks}\ and\ \citenamefont
  {Fairbank}(1968)}]{Wilks1968}%
  \BibitemOpen
  \bibfield  {author} {\bibinfo {author} {\bibfnamefont {J.}~\bibnamefont
  {Wilks}}\ and\ \bibinfo {author} {\bibfnamefont {H.~A.}\ \bibnamefont
  {Fairbank}},\ }\href {\doibase 10.1119/1.1975128} {\bibfield  {journal}
  {\bibinfo  {journal} {Am. J. of Phys.}\ }\textbf {\bibinfo
  {volume} {36}},\ \bibinfo {pages} {764} (\bibinfo {year} {1968})}\BibitemShut
  {NoStop}%
\bibitem [{\citenamefont {Lacey}\ \emph {et~al.}(2007)\citenamefont {Lacey},
  \citenamefont {Ajitanand}, \citenamefont {Alexander}, \citenamefont {Chung},
  \citenamefont {Holzmann}, \citenamefont {Issah}, \citenamefont {Taranenko},
  \citenamefont {Danielewicz},\ and\ \citenamefont {St{\"o}cker}}]{Lacey2007}%
  \BibitemOpen
  \bibfield  {author} {\bibinfo {author} {\bibfnamefont {R.~A.}\ \bibnamefont
  {Lacey}}, \bibinfo {author} {\bibfnamefont {N.~N.}\ \bibnamefont
  {Ajitanand}}, \bibinfo {author} {\bibfnamefont {J.~M.}\ \bibnamefont
  {Alexander}}, \bibinfo {author} {\bibfnamefont {P.}~\bibnamefont {Chung}},
  \bibinfo {author} {\bibfnamefont {W.~G.}\ \bibnamefont {Holzmann}}, \bibinfo
  {author} {\bibfnamefont {M.}~\bibnamefont {Issah}}, \bibinfo {author}
  {\bibfnamefont {A.}~\bibnamefont {Taranenko}}, \bibinfo {author}
  {\bibfnamefont {P.}~\bibnamefont {Danielewicz}}, \ and\ \bibinfo {author}
  {\bibfnamefont {H.}~\bibnamefont {St{\"o}cker}},\ }\href {\doibase
  10.1103/physrevlett.98.092301} {\bibfield  {journal} {\bibinfo  {journal}
  {Phys. Rev. Lett.}\ }\textbf {\bibinfo {volume} {98}},\ \bibinfo {pages}
  {092301} (\bibinfo {year} {2007})}\BibitemShut {NoStop}%
\bibitem [{\citenamefont {Sch{\"a}fer}\ and\ \citenamefont
  {Teaney}(2009)}]{Schaefer2009}%
  \BibitemOpen
  \bibfield  {author} {\bibinfo {author} {\bibfnamefont {T.}~\bibnamefont
  {Sch{\"a}fer}}\ and\ \bibinfo {author} {\bibfnamefont {D.}~\bibnamefont
  {Teaney}},\ }\href {\doibase 10.1088/0034-4885/72/12/126001} {\bibfield
  {journal} {\bibinfo  {journal} {Rep. Prog. Phys.}\ }\textbf
  {\bibinfo {volume} {72}},\ \bibinfo {pages} {126001} (\bibinfo {year}
  {2009})}\BibitemShut {NoStop}%
  %
  \bibitem [{\citenamefont {Bradlyn}\, \citenamefont{Goldstein},\ and\ \citenamefont
  {Read}(2009)}]{Bradlyn2012}%
  \BibitemOpen
  \bibfield  {author} {\bibinfo {author} {\bibfnamefont {B.}~\bibnamefont
  {Bradlyn}},\ \bibinfo {author} {\bibfnamefont {M.}~\bibnamefont
  {Goldstein}},\ and\ \bibinfo {author} {\bibfnamefont {N.}~\bibnamefont
  {Read}},\ }\href {\doibase 10.1103/PhysRevB.86.245309} {\bibfield
  {journal} {\bibinfo  {journal} {Phys. Rev. B}\ }\textbf
  {\bibinfo {volume} {86}},\ \bibinfo {pages} {245309} (\bibinfo {year}
  {2012})}\BibitemShut {NoStop}%
  %
  %
  \bibitem [{\citenamefont {Zaanen}\, \citenamefont{Liu},\ \citenamefont{Sun},\ and \ \citenamefont{Schlam}(2015)}]{Zaanen2015}%
  \BibitemOpen
  \bibfield  {author} {\bibinfo {author} {\bibfnamefont {J.}~\bibnamefont
  {Zaanen}},\ \bibinfo {author} {\bibfnamefont {Y.}~\bibnamefont{Liu}},\ \bibinfo {author} {\bibfnamefont{Y.}~ \bibfnamefont{Sun}}, \ and \ \bibinfo {author} {\bibfnamefont{K.}~ \bibfnamefont{Schlam},\ } }\href {\doibase 10.1017/CBO9781139942492} {\emph {\bibinfo
  {title} {Holographic Duality in Condensed Matter Physics}}}\ (\bibinfo  {publisher}
  {Cambridge University Press},\ \bibinfo {year} {2015})\BibitemShut {NoStop}%
  %
  %
  %
   \bibitem [{\citenamefont {Hartnoll}, and \ \citenamefont{Lucas},\ \citenamefont{Sachdev}(2018)}]{Hartnoll2018}%
  \BibitemOpen
  \bibfield  {author} {\bibinfo {author} {\bibfnamefont {S.}~\bibfnamefont{A.}~\bibnamefont
  {Hartnoll}},\ \bibinfo {author} {\bibfnamefont {A.}~\bibnamefont{Lucas}}, and \ \bibinfo {author} {\bibfnamefont{S.}~ \bibfnamefont{Sachdev}},\ }\href {https://arxiv.org/abs/1612.07324} {\emph {\bibinfo
  {title} {Holographic quantum matter}}}\ (\bibinfo  {publisher}
  {MIT Press},\ \bibinfo {year} {2018})\BibitemShut {NoStop}%
  %
  %
\bibitem [{\citenamefont {Abrikosov}\ and\ \citenamefont
  {Beneslavski{\u{i}}}(1996)}]{ABRIKOSOV1996}%
  \BibitemOpen
  \bibfield  {author} {\bibinfo {author} {\bibfnamefont {A.~A.}\ \bibnamefont
  {Abrikosov}}\ and\ \bibinfo {author} {\bibfnamefont {S.~D.}\ \bibnamefont
  {Beneslavski{\u{i}}}},\ }in\ \href {\doibase 10.1142/9789814317344_0010}
  {\emph {\bibinfo {booktitle} {30 Years of the Landau Institute {\textemdash}
  Selected Papers}}}\ (\bibinfo  {publisher} {{WORLD} {SCIENTIFIC}},\ \bibinfo
  {year} {1996})\ pp.\ \bibinfo {pages} {64--73}\BibitemShut {NoStop}%
\bibitem [{\citenamefont {Janssen}\ and\ \citenamefont
  {Herbut}(2015)}]{Janssen2015}%
  \BibitemOpen
  \bibfield  {author} {\bibinfo {author} {\bibfnamefont {L.}~\bibnamefont
  {Janssen}}\ and\ \bibinfo {author} {\bibfnamefont {I.~F.}\ \bibnamefont
  {Herbut}},\ }\href {\doibase 10.1103/physrevb.92.045117} {\bibfield
  {journal} {\bibinfo  {journal} {Phys. Rev. B}\ }\textbf {\bibinfo {volume}
  {92}},\ \bibinfo {pages} {045117} (\bibinfo {year} {2015})}\BibitemShut
  {NoStop}%
\bibitem [{\citenamefont {Moon}\ \emph {et~al.}(2013)\citenamefont {Moon},
  \citenamefont {Xu}, \citenamefont {Kim},\ and\ \citenamefont
  {Balents}}]{Moon2013}%
  \BibitemOpen
  \bibfield  {author} {\bibinfo {author} {\bibfnamefont {E.-G.}\ \bibnamefont
  {Moon}}, \bibinfo {author} {\bibfnamefont {C.}~\bibnamefont {Xu}}, \bibinfo
  {author} {\bibfnamefont {Y.~B.}\ \bibnamefont {Kim}}, and\ \bibinfo
  {author} {\bibfnamefont {L.}~\bibnamefont {Balents}},\ }\href {\doibase
  10.1103/physrevlett.111.206401} {\bibfield  {journal} {\bibinfo  {journal}
  {Phys. Rev. Lett.}\ }\textbf {\bibinfo {volume} {111}},\ \bibinfo {pages}
  {206401} (\bibinfo {year} {2013})}\BibitemShut {NoStop}%
\bibitem [{\citenamefont {Boettcher}\ and\ \citenamefont
  {Herbut}(2017)}]{Boettcher2017}%
  \BibitemOpen
  \bibfield  {author} {\bibinfo {author} {\bibfnamefont {I.}~\bibnamefont
  {Boettcher}}\ and\ \bibinfo {author} {\bibfnamefont {I.~F.}\ \bibnamefont
  {Herbut}},\ }\href {\doibase 10.1103/physrevb.95.075149} {\bibfield
  {journal} {\bibinfo  {journal} {Phys. Rev. B}\ }\textbf {\bibinfo {volume}
  {95}},\ \bibinfo {pages} {075149} (\bibinfo {year} {2017})}\BibitemShut
  {NoStop}%
\bibitem [{\citenamefont {Herbut}(2007)}]{Herbut2007}%
  \BibitemOpen
  \bibfield  {author} {\bibinfo {author} {\bibfnamefont {I.}~\bibnamefont
  {Herbut}},\ }\href {\doibase 10.1017/cbo9780511755521} {\emph {\bibinfo
  {title} {A Modern Approach to Critical Phenomena}}}\ (\bibinfo  {publisher}
  {Cambridge University Press},\ \bibinfo {year} {2007})\BibitemShut {NoStop}%
\bibitem [{\citenamefont {Herbut}\ and\ \citenamefont
  {Janssen}(2014)}]{Herbut2014}%
  \BibitemOpen
  \bibfield  {author} {\bibinfo {author} {\bibfnamefont {I.~F.}\ \bibnamefont
  {Herbut}}\ and\ \bibinfo {author} {\bibfnamefont {L.}~\bibnamefont
  {Janssen}},\ }\href {\doibase 10.1103/physrevlett.113.106401} {\bibfield
  {journal} {\bibinfo  {journal} {Phys. Rev. Lett.}\ }\textbf {\bibinfo
  {volume} {113}},\ \bibinfo {pages} {106401} (\bibinfo {year}
  {2014})}\BibitemShut {NoStop}%
\bibitem [{\citenamefont {Janssen}\ and\ \citenamefont
  {Herbut}(2017)}]{Janssen2017}%
  \BibitemOpen
  \bibfield  {author} {\bibinfo {author} {\bibfnamefont {L.}~\bibnamefont
  {Janssen}}\ and\ \bibinfo {author} {\bibfnamefont {I.~F.}\ \bibnamefont
  {Herbut}},\ }\href {\doibase 10.1103/physrevb.95.075101} {\bibfield
  {journal} {\bibinfo  {journal} {Phys. Rev. B}\ }\textbf {\bibinfo {volume}
  {95}},\ \bibinfo {pages} {075101} (\bibinfo {year} {2017})}\BibitemShut
  {NoStop}%
\bibitem [{\citenamefont {Danielewicz}(1984)}]{Danielewicz1984}%
  \BibitemOpen
  \bibfield  {author} {\bibinfo {author} {\bibfnamefont {P.}~\bibnamefont
  {Danielewicz}},\ }\href {\doibase 10.1016/0003-4916(84)90092-7} {\bibfield
  {journal} {\bibinfo  {journal} {Ann. Phys.}\ }\textbf {\bibinfo
  {volume} {152}},\ \bibinfo {pages} {239} (\bibinfo {year}
  {1984})}\BibitemShut {NoStop}%
\bibitem [{\citenamefont {Dumitrescu}(2015)}]{PhysRevB.92.121102}%
  \BibitemOpen
  \bibfield  {author} {\bibinfo {author} {\bibfnamefont {P.~T.}\ \bibnamefont
  {Dumitrescu}},\ }\href {\doibase 10.1103/PhysRevB.92.121102} {\bibfield
  {journal} {\bibinfo  {journal} {Phys. Rev. B}\ }\textbf {\bibinfo {volume}
  {92}},\ \bibinfo {pages} {121102(R)} (\bibinfo {year} {2015})}\BibitemShut
  {NoStop}%
\bibitem [{\citenamefont {Link}\ \emph
  {et~al.}(2018{\natexlab{b}})\citenamefont {Link}, \citenamefont {Sheehy},
  \citenamefont {Narozhny},\ and\ \citenamefont {Schmalian}}]{Link2018a}%
  \BibitemOpen
  \bibfield  {author} {\bibinfo {author} {\bibfnamefont {J.~M.}\ \bibnamefont
  {Link}}, \bibinfo {author} {\bibfnamefont {D.~E.}\ \bibnamefont {Sheehy}},
  \bibinfo {author} {\bibfnamefont {B.~N.}\ \bibnamefont {Narozhny}}, \ and\
  \bibinfo {author} {\bibfnamefont {J.}~\bibnamefont {Schmalian}},\ }\href
  {\doibase 10.1103/physrevb.98.195103} {\bibfield  {journal} {\bibinfo
  {journal} {Phys. Rev. B}\ }\textbf {\bibinfo {volume} {98}},\ \bibinfo
  {pages} {195103} (\bibinfo {year} {2018}{\natexlab{b}})}\BibitemShut
  {NoStop}%
\bibitem [{\citenamefont {Mihaila}\ \emph {et~al.}(2017)\citenamefont
  {Mihaila}, \citenamefont {Zerf}, \citenamefont {Ihrig}, \citenamefont
  {Herbut},\ and\ \citenamefont {Scherer}}]{Mihaila2017}%
  \BibitemOpen
  \bibfield  {author} {\bibinfo {author} {\bibfnamefont {L.~N.}\ \bibnamefont
  {Mihaila}}, \bibinfo {author} {\bibfnamefont {N.}~\bibnamefont {Zerf}},
  \bibinfo {author} {\bibfnamefont {B.}~\bibnamefont {Ihrig}}, \bibinfo
  {author} {\bibfnamefont {I.~F.}\ \bibnamefont {Herbut}}, \ and\ \bibinfo
  {author} {\bibfnamefont {M.~M.}\ \bibnamefont {Scherer}},\ }\href {\doibase
  10.1103/physrevb.96.165133} {\bibfield  {journal} {\bibinfo  {journal} {Phys.
  Rev. B}\ }\textbf {\bibinfo {volume} {96}},\ \bibinfo {pages} {165133}
  (\bibinfo {year} {2017})}\BibitemShut {NoStop}%
\bibitem [{\citenamefont {Ihrig}\ \emph {et~al.}(2019)\citenamefont {Ihrig},
  \citenamefont {Zerf}, \citenamefont {Marquard}, \citenamefont {Herbut},\ and\
  \citenamefont {Scherer}}]{Ihrig2019}%
  \BibitemOpen
  \bibfield  {author} {\bibinfo {author} {\bibfnamefont {B.}~\bibnamefont
  {Ihrig}}, \bibinfo {author} {\bibfnamefont {N.}~\bibnamefont {Zerf}},
  \bibinfo {author} {\bibfnamefont {P.}~\bibnamefont {Marquard}}, \bibinfo
  {author} {\bibfnamefont {I.~F.}\ \bibnamefont {Herbut}}, \ and\ \bibinfo
  {author} {\bibfnamefont {M.~M.}\ \bibnamefont {Scherer}},\ }\href {\doibase
  10.1103/physrevb.100.134507} {\bibfield  {journal} {\bibinfo  {journal}
  {Phys. Rev. B}\ }\textbf {\bibinfo {volume} {100}},\ \bibinfo {pages}
  {134507} (\bibinfo {year} {2019})}\BibitemShut {NoStop}%
  %
\end{thebibliography}
%
%
%
\end{document}